%% file: main.tex
\def \<{\langle}
\def \>{\rangle}
 
\documentclass[a4paper,11pt]{article}  %>>> use for US letter paper
\usepackage{jcappub,graphicx,epsfig,natbib,color,times,bm,amsmath,multirow,hyperref,enumerate}
\usepackage{amsthm,amssymb}

\usepackage{mathrsfs}
\usepackage[flushleft]{threeparttable}

\title{Study on the filters of atmospheric contamination in ground based CMB observation}

\input{authors.tex}

\begin{document}

\abstract{
\input{abstract.tex}

}

%\keywords{methods: data analysis – CMB polarization $B$ mode}
\maketitle

\section{Introduction}\label{sec:intro}
\input{introduction.tex}

%\section{Science motivation of calibration}\label{sec:metho}
%\input{objective.tex}

\section{Atmosphere effects}\label{sec:atmosphere-rev}
\input{atm-review.tex}

\section{Filters}\label{sec:metho}
\input{filter-intro.tex}

\section{Data Simulations and processing}\label{sec:simu}
\input{simulation.tex}

\section{Analysis and results}\label{sec:result}
\input{results.tex}

\section{Summary and discussion}%\label{sec:conclusions}
\input{conclusion.tex}

\acknowledgments
We thank Dr. Zi-Rui Zhang, Zi-Xuan Zhang, and Yi-Ming Wang, Hua Zhai for useful discussion. This study is supported in part by the National Key R\&D Program of China No.2020YFC2201600 and No. 2021YFC2203100, and by the NSFC No.11653002.

\input{appendix}

\nocite{*}
\bibliography{main}
\bibliographystyle{JHEP} 

\end{document}

%% file: authors.tex
\author[a]{Yi-Wen Wu,}
\author[a]{SiYu Li\footnote{Corresponding author\label{corresauth}},}
\author[a]{Yang Liu,}
\author[c]{Zirui Zhang,}
\author[b,a]{Hao Liu,}
\author[a]{Hong Li\textsuperscript{\ref{corresauth}}}\emailAdd{hongli@ihep.ac.cn}%\emailAdd{hongli@ihep.ac.cn}

\affiliation[a]{Key Laboratory of Particle Astrophysics, Institute of High Energy Physics, Chinese Academy of Sciences, 19B Yuquan Road, Beijing 100049, People’s Republic of China}
\affiliation[b]{School of Physics and Optoelectronics Engineering, Anhui University, 111 Jiulong Road, Hefei, Anhui, 230601, People’s Republic of China}
\affiliation[c]{Institute of Frontier and Interdisciplinary Science and Key Laboratory of Particle Physics and Particle Irradiation (MOE), Shandong University, Qingdao}

%% file: abstract.tex
The atmosphere is one of the most significant sources of contamination in ground-based Cosmic Microwave Background (CMB) observations. Atmospheric emission increases the additional optical loading on the detector, resulting in higher photon noise. Additionally, atmospheric fluctuations cause spatial and temporal variations in detected power, leading to additional correlations between detectors and in the time stream of individual detectors. This correlated signal, known as the $1/f$ noise, can interfere with the detection of CMB signals, severely hindering the probing of CMB signals. 
In this paper, we study three types of filters: the polynomial fitting, high-pass filter, and Wiener filter. We evaluate the filters based on their ability to remove atmospheric noise, and investigate the impact of the filters on the data analytic process through end-to-end simulations of CMB experiments. We track their performance by analyzing the response of different components of the data, including both signal and noise. In the time domain, the high-pass filter is found to have the smallest root mean square error and achieves high filtering efficiency, followed by the Wiener filter and polynomial fitting. We adopt two map making methods, namely naive map making and Minimum Variance map making, to study the effects of filters on the map level. The results show that the polynomial fitting gives a high noise residual at low frequency, resulting in significant leakage to small scales in the map domain, while the high-pass and Wiener filters do not have significant leakage.
We compare the filters' effects on the power spectra domain by estimating the angular power spectra of residual noise and input signal, and estimating the standard deviation of the signal recovered power spectra. At low noise levels, the three filters give almost comparable standard deviations on medium and small scales. However, at high noise level, the standard deviation of the polynomial fitting is significantly larger. These studies can be used for reducing atmospheric noise in future ground-based CMB data processing.

%% file: introduction.tex
Cosmic microwave background radiation (CMB), the legacy radiation from the cosmic inflation that is commonly believed to have occurred at the  beginning of universe. This radiation has been a powerful cosmological probe since it was first discovered by Penzias and Wilson over 50 years ago\cite{1965ApJ...142..419P}.  The temperature and polarization inhomogeneities of CMB photons carry valuable information that can be used to trace back to the early moments of our universe. These inhomogeneities were seeded by the quantum fluctuations of space and time during inflation, and the observation of CMB fluctuations across the sky has always been a focal point of cosmological research.

Over the past 30 years, numerous experiments, including both space missions and ground telescopes, have been conducted to measure the temperature and polarization fluctuations of CMB with increasing accuracy \cite{2007ApJS..170..288H,2020A&A...641A...1P,2014ApJ...792...62B}. The latest observations from the Planck satellite have provided the most precise measurement of the anisotropic power spectra of CMB temperature and polarization, leading to the current best-fit $\rm \Lambda CDM$ model and determining the six major cosmological parameters with uncertainties of just a few percent \cite{2020A&A...641A...5P}. In addition, there are many ongoing CMB telescopes on the ground, including the South Pole Telescope \cite{2011PASP..123..568C} located at the Antarctic pole, the Atacama Cosmology Telescope \cite{2010SPIE.7741E..1SN}, and POLARBEAR \cite{2014ApJ...794..171P} located at the Atacama Desert Observatory in Chile, which provide complementary and comparable results to space observations.

Ground-based CMB telescopes offer several advantages over space missions, which have made them popular in the field. One significant advantage is the ability to achieve large aperture, which enables high-resolution observations of the CMB, providing a powerful complement to space missions. For example, the 10m SPT telescope has achieved arcminute-scale resolution observations of the CMB\cite{2011PASP..123..568C}. In contrast, the diffraction limit of the telescope main-dish aperture restricts the angular resolution of CMB maps observed by space missions, such as COBE, WMAP, and Planck, to about $7^\circ$, $15^\prime$, and $5^\prime$, respectively\cite{1996ApJ...470..653K,2003ApJ...583....1B,2006astro.ph..4069T}. Moreover, small-aperture ground-based telescopes, such as BICEP/KECK\cite{2015ApJ...806..206B,2014SPIE.9153E..1NA} located in Antarctica, offer superior observation performance. In the northern hemisphere, the 72cm AliCPT telescope\cite{Li:2017drr,Li:2018rwc,Salatino_2020,2021ITAS...3165289S,Ghosh:2022mje} is under construction and will be deployed at a site 5250m above sea level in the Ngari region of Tibet to detect primordial gravitational waves, providing a good complement to existing observations in the Southern Hemisphere. Furthermore, ground-based telescopes are more cost-effective than space missions.

Futhermore, the development of superconducting transition edge sensors (TES) and integration technology has opened up new avenues for further sensitivity improvement in ground-based telescopes. The third-generation ground-based telescopes, which have implemented thousands of detectors, have achieved remarkable results. For instance, BICEP3 has a total of 2560 detectors, with an accuracy of about $10\mu$K$\sqrt{s}$\cite{2014SPIE.9153E..1NA}. The fourth-generation telescopes are expected to deploy tens to hundreds of thousands of multi-channel detectors in the focal plane, enabling unprecedented sensitivity levels. The Simons Observatory\cite{2019BAAS...51g.147L}, located in the Atacama Desert in the southern hemisphere, is currently under construction. CMB-S4\cite{2019arXiv190704473A} is in the planning stage and aims to achieve a sensitivity of $1\mu$K$\cdot$ arcmin and a map depth of $\Delta T$ over a few percent of the sky patch in the next decade.

However, Ground-based observations of the CMB are subject to interference from microwave absorption and emission due to the presence of the Earth's atmosphere. The atmosphere causes microwave attenuation, which severely reduces the transmission of CMB photons, limiting ground-based observations to a few microwave windows. Additionally, the absorption of certain bands of photons leads to atmospheric background emission, causing severe noise interference to CMB observations. Even at high-altitude sites where the atmosphere is very thin and dry, the background brightness temperature contributed by atmospheric radiation can reach about $10$K or even higher. Moreover, atmospheric variations, such as turbulence, lead to fluctuations in atmospheric radiation over time and space, creating patterns in the sky map that contaminate measurements of CMB anisotropy. To suppress this interference, filters are required in the time domain. In this paper, we focus on studying the effectiveness of different types of filters, which are polynomial fitting, high-pass and Wiener filter.

The paper is organized as follows: In Section \ref{sec:atm_contamination}, we introduce the effects of atmospheric contamination on microwave observations. Section \ref{sec:filters} presents the filtering methods. In Section \ref{sec:simulation}, we describe the data simulation process, and in Section \ref{sec:results}, we present the results and discussion. Finally, we summarize our conclusions and discuss their implications in Section \ref{sec:conclusions}.

%% file: atm-review.tex
\label{sec:atm_contamination}
\subsection{Emission and absorption}
Microwave band observations are affected by the atmosphere for two main reasons. Firstly, certain components of the atmosphere such as oxygen and water vapor molecules absorb microwave radiation. Secondly, the movement of these molecules generates additional emission. Oxygen molecules have two absorption bands in the microwave range with wavelengths of 4-6 mm (frequency of 46.45-71.05 GHz) and 2.53 mm (frequency of 118.75 GHz). Water vapor molecules also have two absorption bands in the microwave range with absorption spectral lines of 1.35 cm (frequency of 22.235 GHz) and 1.64 mm (frequency of 183 GHz). Therefore, the ground-based CMB experiments typically operate at frequencies below 50 GHz, near 100 GHz, near 220 GHz, and near 270 GHz.

CMB telescopes are often built in the driest places on earth to minimize atmospheric interference. There are currently four established sites, including the South Pole in Antarctica (2835 m) and the Chajnantor Plateau in Chile (4990 m) located in the southern hemisphere, as well as the Summit Station in Greenland (3216 m) and the Ali observatory in Tibet (5250 m) located in the northern hemisphere. These sites generally have very low Precipitable Water Vapor (PWV) values, with median values usually below 1 mm measured by MERRA-2 during the observation season.\cite{2017arXiv170909053L,2017ApJ...848...64K}.

Secondly, the strong emission from the atmosphere also poses a significant challenge for ground-based CMB observations. It introduces an additional term $E(\nu)$ to the total background emission detected. The atmospheric temperature near the Earth's surface is approximately 280 K \cite{2015ApJ...809...63E}, which contributes to a total background emission of around 20 K at a frequency of 150 GHz. The additional term $E(\nu)$ can be expressed as follows:
\begin{equation}
E(\nu)= [1-T(\nu)] B_{\nu}(T_{\mathrm{atm}}),
\end{equation}
where $T_{\mathrm{atm}} \sim 280,\mathrm{K}$ is the atmospheric temperature, $B_{\nu}(T_{\mathrm{atm}})$ is the thermal equilibrium blackbody spectrum at atmospheric temperature $T_{\mathrm{atm}}$, and $T(\nu)$ is the transmittance of the atmosphere at frequency $\nu$. For the detector, the atmospheric background emission primarily manifests as white noise.

\subsection{Fluctuation}
In addition to the photon noise induced by the atmospheric background emission, which is directly related to the Precipitable Water Vapor (PWV) level, the non-uniformity of the emission caused by atmospheric fluctuations (such as clouds and turbulence) affects the detection of CMB photons. The turbulence of the atmospheric emission along the line of sight is recorded in the detector's time stream readings and results in a strong correlation between the time streams of adjacent detector readings. This signal of atmospheric fluctuation is predominantly distributed at low frequencies, where its power spectral density is inversely proportional to frequency and is often referred to as $1/f$ noise\cite{2020MNRAS.491.4254C,2016arXiv160609584H}. Unlike white noise, this scale-dependent noise contamination cannot be averaged out by multiple measurements or stacking and needs to be filtered out. For the ACT experiment, the median atmospheric temperature drift is roughly distributed at $220$ mK over a 15-minute observation, with a PWV value below 1 mm on a good observing night\cite{2013ApJ...762...10D}. 

Atmospheric emission fluctuations are correlated at different spatial scales\cite{2018RPPh...81d4901S}. The structure of atmospheric turbulence can be modeled by the Kolmogorov model, which is given by the formula $P(k)\propto k^m$, where $k$ is the modes of atmospheric turbulence, and $m=-\frac{11}{3}$\cite{2005ApJ...622.1343B,2000ApJ...543..787L}. Based on the Kolmogorov model, many articles have studied the correlation of atmospheric emission between different spatial modes. Church et al.\cite{1995MNRAS.272..551C} discussed the influence of wind speed and the typical scale of turbulence on the power spectral density (PSD) of atmospheric emission fluctuations, based on a three-dimensional model. Lay \& Halverson\cite{2000ApJ...543..787L} established a two-dimensional atmospheric model. Building on Church's research, Errard\cite{2015ApJ...809...63E} studied how scanning strategies and atmospheric conditions affect atmospheric correlation in detector data streams and derived the correlation function that quantitatively describes this effect. These models have been verified in real experimental observations. For example, the ACT experiment\cite{2013ApJ...762...10D} gives the relationship between the power spectrum of Time Ordered Data (TOD) and the frequency in the form of a power law: $P(f)\sim f^{-\beta}$, where $f$ is frequency and $\beta=1\sim 3.5$. In the ACT experiment, the higher the PWV value of the atmosphere, the larger the value of $\beta$, which is consistent with the model.

The correlation property of atmosphere emission fluctuation indicates that the atmosphere-induced noise at low frequencies cannot be removed effectively through accumulating the data of different detectors, and specific filters are needed in data processing.

%% file: filter-intro.tex
\label{sec:filters}
In this section, we introduce three commonly used filters, which are polynomial fitting, high-pass filter, and Wiener filter. We describe their filtering principles and implementation respectively.

\subsection{Polynomial fitting }
%Polynomial
Polynomial fitting is a widely used method in CMB ground-based experiments\cite{2018ApJ...852...97H,2019ApJ...870..102T,2010ApJ...711.1123C,2011ApJ...741...81B,2013ApJ...765...64M}. With the fitting method of least squares, we can determine the optimal fitting curve by minimizing the sum of squares of errors between the fitting curve and the data. Given a data set $(x_i,y_i ),i=1,2,\ldots,m,$ fitting it with a polynomial $P(x_i)=a_0+a_1 x_i+a_2 x_i^2+\dotsb+a_nx_i^n$ of order $n$, we can derive a linear function group:
\begin{equation}
\begin{cases}
	\begin{array}{l}
	P_n\left( x_1 \right) =a_0+a_1x_1+a_2x_{1}^{2}+...+a_nx_{1}^{n}\\
	P_n\left( x_2 \right) =a_0+a_1x_2+a_2x_{2}^{2}+...+a_nx_{2}^{n}\\
\end{array}\\
	\,\,     \vdots\\
	P_n\left( x_m \right) =a_0+a_1x_m+a_2x_{m}^{2}+...+a_nx_{m}^{n}.\\
\end{cases}
\end{equation}
The sum of squares of errors between the polynomial and the data set is:
\begin{equation}
J=\sum_{i=1}^m{\left[ P\left( x_i \right) -y_i \right] ^2}.
\end{equation}
Written in the form of matrices, it follows:
\begin{equation}
\begin{aligned}
J&=\left( \left[ \begin{matrix}
	1&		x_1&		x_{1}^{2}&		...&		x_{1}^{n}\\
	1&		x_2&		x_{2}^{2}&		...&		x_{2}^{n}\\
	1&		x_3&		x_{3}^{2}&		...&		x_{3}^{n}\\
	\vdots&		\vdots&		\vdots&		\ddots&		\vdots\\
	1&		x_m&		x_{m}^{2}&		...&		x_{m}^{n}\\
\end{matrix} \right] \cdot \left[ \begin{array}{l}
	a_0\\
	a_1\\
	a_2\\
	\vdots\\
	a_n\\
\end{array} \right] -\left[ \begin{array}{l}
	y_0\\
	y_1\\
	y_2\\
	\vdots\\
	y_n\\
\end{array} \right] \right) ^T\cdot \left( \left[ \begin{matrix}
	1&		x_1&		x_{1}^{2}&		...&		x_{1}^{n}\\
	1&		x_2&		x_{2}^{2}&		...&		x_{2}^{n}\\
	1&		x_3&		x_{3}^{2}&		...&		x_{3}^{n}\\
	\vdots&		\vdots&		\vdots&		\ddots&		\vdots\\
	1&		x_m&		x_{m}^{2}&		...&		x_{m}^{n}\\
\end{matrix} \right] \cdot \left[ \begin{array}{l}
	a_0\\
	a_1\\
	a_2\\
	\vdots\\
	a_n\\
\end{array} \right] -\left[ \begin{array}{l}
	y_0\\
	y_1\\
	y_2\\
	\vdots\\
	y_n\\
\end{array} \right] \right) 
\\
&=\left( Xa-Y \right) ^T\left( Xa-Y \right),
\end{aligned}
\end{equation}
where
\begin{equation}
X=\left[ \begin{matrix}
	1&		x_1&		x_{1}^{2}&		...&		x_{1}^{n}\\
	1&		x_2&		x_{2}^{2}&		...&		x_{2}^{n}\\
	1&		x_3&		x_{3}^{2}&		...&		x_{3}^{n}\\
	\vdots&		\vdots&		\vdots&		\ddots&		\vdots\\
	1&		x_m&		x_{m}^{2}&		...&		x_{m}^{n}\\
\end{matrix} \right] ,\;\;\;a=\left[ \begin{array}{l}
	a_0\\
	a_1\\
	a_2\\
	\vdots\\
	a_n\\
\end{array} \right] ,\;\;\;Y=\left[ \begin{array}{l}
	y_0\\
	y_1\\
	y_2\\
	\vdots\\
	y_n\\
\end{array} \right] ,
\end{equation}
and $X$ is known as the Vandermonde matrix. The final equation to estimate the coefficients which minimize the total error can be written as:
\begin{equation}\label{EQ_polyfit}
\begin{aligned}
	\frac{\partial J}{\partial a}&=\frac{\partial \left[ \left( Xa-Y \right) ^T\left( Xa-Y \right) \right]}{\partial a}\\
	&=X^TXa-X^TY\\
	&=0.\\
\end{aligned}
\end{equation}
We can derive the polynomial coefficients:
\begin{equation}
a=\left( X^TX \right) ^{-1}X^TY.
\end{equation}
When applying polynomial fitting into CMB data processing procedure, we should first determine the proper order of polynomial, then calculate the polynomial coefficients to get the fitting curve, and finally subtract it from the TOD to get the filtered data.

\subsection{High-pass filter}
High-pass filter performs noise removal in frequency space by cutting off the TOD on a specific frequency range which is considered to be dominated by noise. Here the cut-off frequency needs to be set to define the frequency limits for passing and blocking. The extent to which low frequencies are attenuated depends on the design of the filter.
The function of the ideal high-pass filter is a step function, which has the following form:
$$
    w(f)=\left\{
                \begin{array}{ll}
                  0 \;\;\;f<f_{cut-off}\\  
                  1 \;\;\;f>f_{cut-off}
                \end{array},
              \right.
$$

    High-pass filter is suitable for ground-based CMB observation time streams to filter out the noise from atmosphere. We know that, atmospheric emission dominates the loading of the detectors, and it is difficult to remove the atmospheric emission from the data in the time domain, especially when the signal-to-noise ratio is very low. However, in the frequency space, since atmospheric noise is mainly concentrated in the low frequency range, we can easily achieve the effect of reducing atmospheric noise through a high-pass filter.
    % In  Section 5.1, we apply a high-pass Butterworth filter on TOD to suppress atmospheric noise.

\subsection{Wiener filter}\label{wiener}
%Wiener filter
Among linear filters using statistical information from data, the generalized Wiener filter is widely used in data processing as a maximum posterior solution for signal and noise statistics\cite{2013A&A...549A.111E,2019MNRAS.490..947K}. 
We discuss the Wiener filter solution of this problem in frequency domain.

Given an input $Y(f)$ at a frequency of $f$, after the filtering operation of $W(f)$, the output signal can be expressed as $\hat{X}(f)=W(f)Y(f)$. The error, which defines the difference between the input signal and the filtered output value, is often a key parameter used to describe the efficiency of filtering, and the error can be writen as
\begin{equation}
\begin{aligned}
\mathscr{E}=X(f)-\hat{X}(f)=X(f)-W(f)Y(f).
\end{aligned}
\end{equation}
The smaller the error, the better the filtering effect, so by finding the minimum value of the error, we will get the best filtering of the noise. We can define the Mean Square of the error (MSE):
\begin{footnotesize}%{small}
\begin{equation}
\begin{aligned}
\mathbb{E}[|\mathscr{E}(f)|^2]&=\mathbb{E}[(X(f)-W(f)Y(f))^*(X(f)-W(f)Y(f))]\\
&=\mathbb{E}[|(X(f)|^2]-W(f)\mathbb{E}[X^*(f)Y(f)]-W(f)^*\mathbb{E}[Y^*(f)X(f)]+|W(f)|^2\mathbb{E}[|Y(f)|^2 ]\\
&=P_{XX}-W(f)P_{XY}-W^*(f)P_{YX}+|W(f)|^2P_{YY}.
\end{aligned}
\end{equation}
\end{footnotesize}

To minimize the MSE value, we can take the derivative of the above equation with respect to $W(f)$ and take it equal to $0$:
\begin{equation}
 \begin{aligned}
\frac{\partial \mathbb{E}[|\mathscr{E}(f)^2|]}{\partial W(f)}=W(f)P_{YY}(f)-P_{XY}(f)=0,
\end{aligned}
\end{equation}
where $P_{YY}(f)=\mathbb{E}[Y(f)Y^*(f)]$ and $P_{XY}(f)=\mathbb{E}[X(f)Y^*(f)]$ are the PSD of input signal $Y(f)$ and the cross-PSD of $Y(f)$ and $X(f)$, respectively. From above equation, we finally get the Wiener filter with minimum MSE as:
\begin{equation}
\begin{aligned}
W(f)=(P_{XY}(f))/(P_{YY}(f)) 
\end{aligned}   
\end{equation}
Since there's no inherent correlation between atmospheric emission and the CMB signal, we can get the following relation:
\begin{equation}
\begin{aligned}
P_{XY}(f)=P_{XX}(f)
P_{YY}(f)=P_{XX}(f)+P_{NN}(f), 
\end{aligned} 
\end{equation}
where $P_{XX}(f)$ and $P_{NN}(f)$ are the PSD of desired signal $X(f)$ and atmospheric emission $N(f)$. The final formula of Wiener solution can be written as:
\begin{align}
W(f) &= P_{XX}(f)/(P_{XX}(f)+P_{NN}(f)) \\ \nonumber
&= 1/(1+\rho^{-1}(f)),
\end{align}\label{wiener solution}
where $\rho(f) = P_{XX}(f)/P_{NN}(f)$ is the signal-to-noise ratio. Therefore, the estimated value after Wiener filtering is:
\begin{equation}
\begin{aligned}
\hat X(f)=W(f)Y(f)=P_{XX}(f)/(P_{XX}(f)+P_{NN}(f))Y(f),
\end{aligned}\label{eq:wf}
\end{equation}

where $P_{XX}(f)$ is estimated from TOD without the atmospheric fluctuation, and $P_{XX}(f)+P_{NN}(f)$ is estimated from total TOD itself.

%% file: simulation.tex
\label{sec:simulation}
In this section, we describe in detail the process of generating the simulated data. We simulate sky signals that are very relevant to the CMB experiment, including the CMB, the major component of galactic foreground radiation from the Milky Way in microwave, like synchrotron, thermal dust, and free-free, and we also simulate atmospheric emission in the microwave band. We then generate time series of detectors readings based on a typical ground-based CMB scanning strategy, provide simulated time-series data stream, and carry out filtering analysis on the time-streams.

\subsection{Sky simulation in microwave band}
For the simulation of CMB sky, we generate a HEALPix\cite{2005ApJ...622..759G} realization of the full sky CMB maps from the angular power spectra obtained by CAMB\cite{Lewis:1999bs}, and the input cosmology is the best fit $\Lambda$CDM model of Planck 2018 data release\cite{2020A&A...641A...5P}.

For foreground emission simulation, we only consider the Gaussian part, and start from the measured angular power spectra of various foreground components, such as synchrotron, dust, free-free, etc., given by the Planck satellite measurements\cite{2015A&A...576A.107P}, and use HEALPix to convert the angular power spectra to the sky maps for each components.

The spectra of free-free, synchrotron and thermal dust follows:
\begin{equation}
    \begin{aligned}
C_{\ell}^{FF}&=0.068\times \left( \frac{f_{sky}}{0.6} \right) ^{\left[ 6.10+3.90\ln\left( f_{sky}/0.6 \right) \right]}\times \left( \frac{\ell}{100} \right) ^{-2.2}\times \left( \frac{\nu}{\nu_b} \right) ^{-4.28},
\\
C_{\ell}^{sync}&=2.96\times 10^9 \times \left( \frac{f_{sky}}{0.6} \right) ^{\left[ 2.12+2.67\ln\left( f_{sky}/0.6 \right) \right]}\times \left( \frac{\ell}{100} \right) ^{-2.5}\times \left( \frac{\nu}{\nu_c} \right) ^{-6.0},
\\
C_{\ell}^{dust}&=0.086\times \left( \frac{f_{sky}}{0.6} \right) ^{\left[ 4.60+7.11\ln\left( f_{sky}/0.6 \right) \right]}\times \left( \frac{\ell}{100} \right) ^{-2.4}\times D_{\nu},
    \end{aligned}
\end{equation}
where $\rm \nu_b=23GHz$, $\rm \nu_c=0.408GHz$, $f_{sky}$ represents the covered sky area of scanning, and $\rm D_\nu$ is a spectral model of the dust emission.

\subsection{Atmospheric emission}\label{sec:atm_emis}

Atmospheric emission contaminates observations in the microwave band and is a major source of noise in the observed data. Based on the characteristics of the atmospheric radiation, we generated $1/f$ noise and white noise respectively, and then combined them in the TOD.

\begin{figure}[bthp]
    \begin{center}
        \includegraphics[width=1\textwidth]{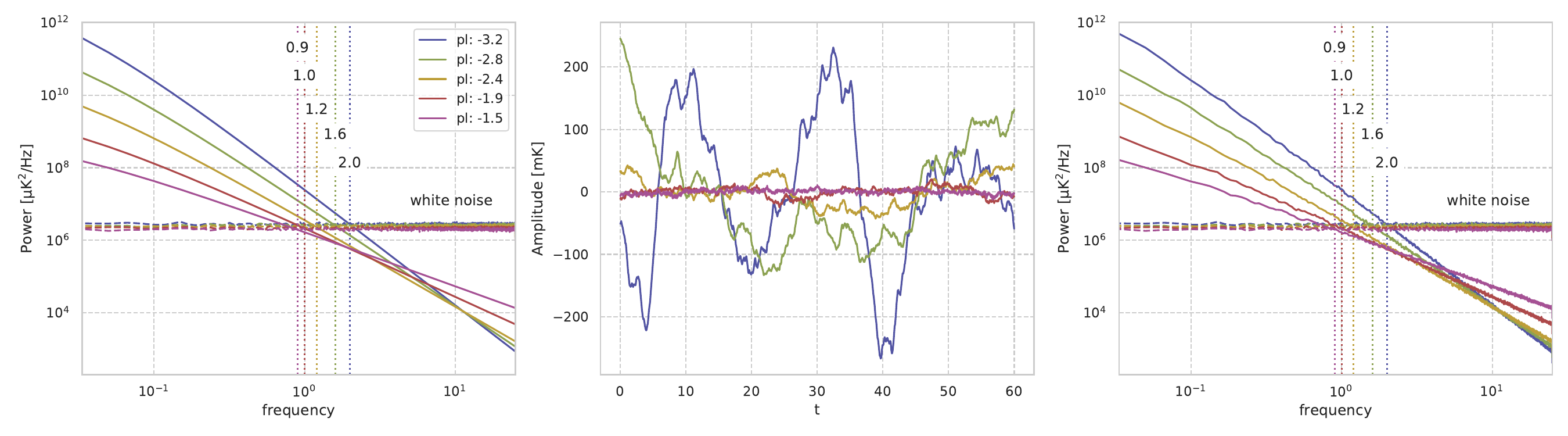}
    \end{center}
    \caption{Left: PSD of observed data from ACT experiments. Middle: the generated TOD from PSD in the left pannel. Right: the PSD of generated TOD. The simulated TOD are consistant with the observation of ACT.}\label{fig:psd}
\end{figure}
\begin{itemize}
    \item $1/f$ noise: This is achieved by the power spectrum $P(f)=\alpha f^{-\beta}$, where $\alpha$ and $\beta$ denote the fluctuation amplitude and the index of the spectral line, respectively. For the determination of these two parameters, we refer to the observed data from ACT experiments to ensure that the noise level and emission fluctuations are as reasonable and close to the real situation as possible\cite{2013ApJ...762...10D}. In Figure \ref{fig:psd}, the left panel is the PSD of observed data from ACT experiments,  and for different power-law indices, we label them in five different colors. The vertical dotted lines represent the knee frequencies at which the atmospheric noise PSD and the white noise PSD intersect with each other for different power-law indices, respectively. The middle panel is the results of generated TOD for different power laws of PSD. In order not to lose generality, we have simulated two levels of $1/f$ noise with both low and high radiation intensities separately, of which the specific PSD parameters of amplitude and spectral index are listed in Table \ref{table:noise}.
    \item White noise: We generated white Gaussian noise with random numbers and, here, we also generated white noise with two levels of intensity, whose parameter selection is listed in Table \ref{table:noise}.
\end{itemize}

\begin{table}[ht!]
    \centering
    \caption{Parameters of noise spectra}
    \begin{tabular}{c|c|c|c}
    \hline
    Noise level & $\beta$  & Knee freq./Hz & White noise level  \\
    \hline
    Low & $-1.5$ & $0.9$ & $1000 \mu K \sqrt{s}$  \\
    \hline
    High & $-3.2$ & $2.0$ & $1400 \mu K \sqrt{s}$  \\
    \hline
    \end{tabular}
    \label{table:noise}
\end{table}

Based on the above simulations of $1/f$ and white noise, we combine these two components to produce low and high intensity atmospheric emission, as defined in the table for the ``low" and ``high" noise level in Table \ref{table:noise}. In subsequent analysis, reference to ``low" or ``high" noise level indicates the presence of both $1/f$ and white noise components.

\subsection{TOD generation}\label{sec:TODgeneration}

%\textcolor{red}{LISY:start}

We simulate an virtual CMB observation with a telescope of 2000 detectors\footnote{ We adopt an ideal model that the detectors are located at the center of its focal plane  and do not include a real beam size for simplicity.},  and generate the TOD data. 
A scan strategy is required to get the trajectory of each detector's pointing over the pre-computed sky maps.
The basic scanning pattern of the telescope is to fix the elevation angle and to carry out a 360 degree azimuth scan.
The values of basic parameters such as elevation angle, scanning speed and sampling frequency are listed in Table \ref{tab:my_label}.
We assume that during the mission the telescope observes 400 days and works for 7 hours per day. We take the spinning of the Earth into account, and for simplicity we ignore the orbiting so that daily scans perfectly overlap with each other.
It is convenient to name each 360 degrees' scan as a Ring scan.
In total each dataset includes 400 days' data, each day has 420 ring scans, and each ring lasts 60 seconds and contains 3,000 samples according to the scan strategy.

\begin{table}[h!]
    \centering
    \caption{Major parameters of scan strategy taken in the simulation}
    \begin{tabular}{c|c|c|c|c}
    \hline
    Site latitude & Elevation & Azimuth range & Scanning speed & Sampling frequency \\
    \hline
     $\rm 32^\circ N$ & $50^{\circ}$ & $360^{\circ}$ & $\rm 6^{\circ}/s$ & $\rm 50Hz$  \\
    \hline
    \end{tabular}
    \label{tab:my_label}
\end{table}

We get the signal time streams by sampling the CMB maps as well as the foreground maps. 
     To generate the TOD data of atmospheric $1/f$ noise and white noise, we apply an inverse Fast Fourier Transform (iFFT) to the square root of their power spectrum density, before which the following procedure are taken:
\begin{itemize}
    \item Convert the power spectrum density to amplitude: $A(f) = \sqrt{2P(f)} = \sqrt{2\alpha f^{-\beta}}$, where factor $2$ represents double sideband PSD, then we get the sample amplitude $\rm A(f)$ in frequency domain.
    \item Generate Gaussian white noise time series $n(t)$: the length of the white noise series is set to the same with samples, and the variance of white noise should be set to one.
    \item Apply FFT to the white noise: $ N(f) = FFT(n(t))$. This step is to acquire random variables with uniform phase distributed between $(0, 2\pi)$.
    \item Multiply the sample amplitude: $S(f) = A(f)*N(f)$. The spectrum of sample frequency components can be generated by multiplying sample amplitude spectrum $ A(f)$ and phase spectrum
    $N(f)$. We set the variance of the generated white noise equal to 1, which guarantees $|N(f)|=1$.
    \item Apply iFFT to the frequency domain representation: $S(t) = iFFT(S(f))$. Note that in actual computation we apply irFFT to $S(f)$ in stead of iFFT by using python package Numpy, as the samples are purely real input, and the output should be Hermitian-symmetric, and finally we take the real part of the result as the generated 
    $1/f$ noise time series.
\end{itemize}

\begin{figure}[bthp]
    \begin{center}
        \includegraphics[width=0.9\textwidth]{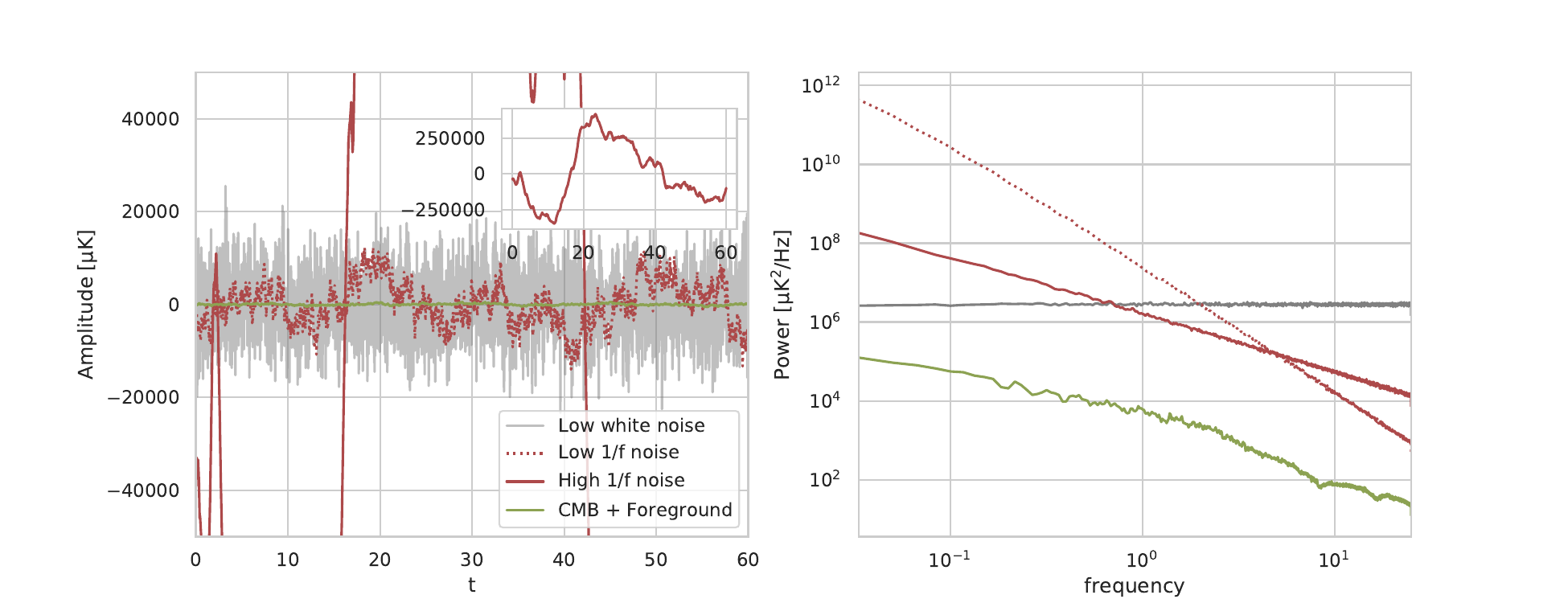}
    \end{center}
    \caption{The generated TOD of four components: CMB and foreground (green line), atmosphere noise with low level (red dotted line ), atmosphere noise with high level (red solid line), white noise with low level (gray line).}\label{fig:tod}
\end{figure}

In Figure \ref{fig:tod}, the left panel shows the simulated TOD while the right panel displays the corresponding power spectral density. The different emission components are distinguished by their respective colored lines. The green line corresponds to the sum of CMB and foreground emissions, and the grey line is for showing the low level white noise at $1000 \mu K\sqrt{s}$. The contributions of atmospheric $1/f$ noise are plotted using both the low level (red dotted line) and high level (red solid line) representations.

\subsection{Filtering operation}
In this section we describe the specific implementation of the filtering on the TOD data.

\subsubsection{Polynomial fitting on TOD}
\begin{figure}[bthp]
    \begin{center}
        \includegraphics[width=1\textwidth]{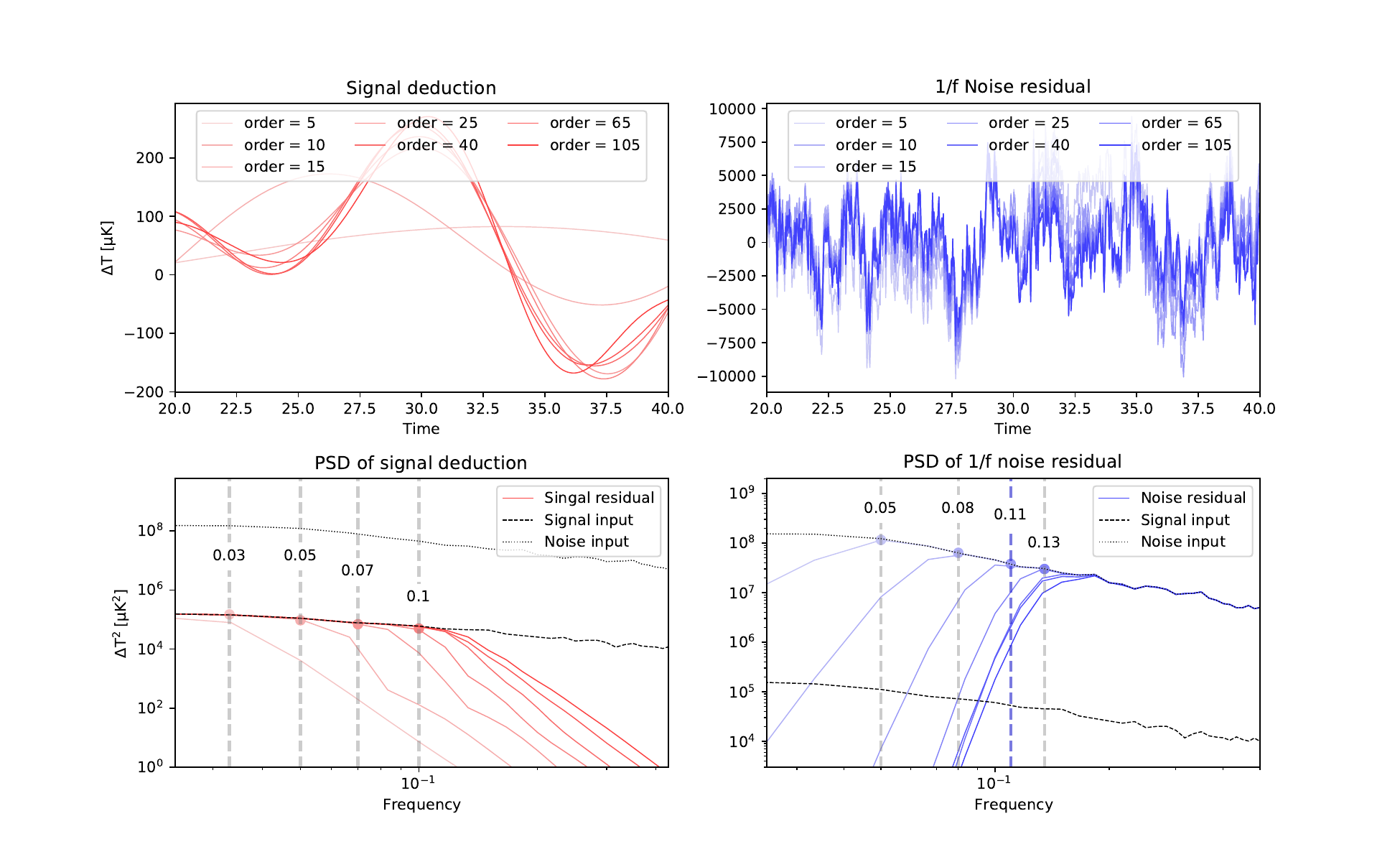}
    \end{center}
    \caption{The order test for low noise level TOD. Upper panel is the fitting curves for signal and $1/f$ noise under order from 5 to 105; lower panel is the corresponding PSD for both components. The blue vertical dashed line labels the corner frequency of 0.11.}\label{fig:poly_n_low}
\end{figure}

\begin{figure}[bthp]
    \begin{center}
        \includegraphics[width=1\textwidth]{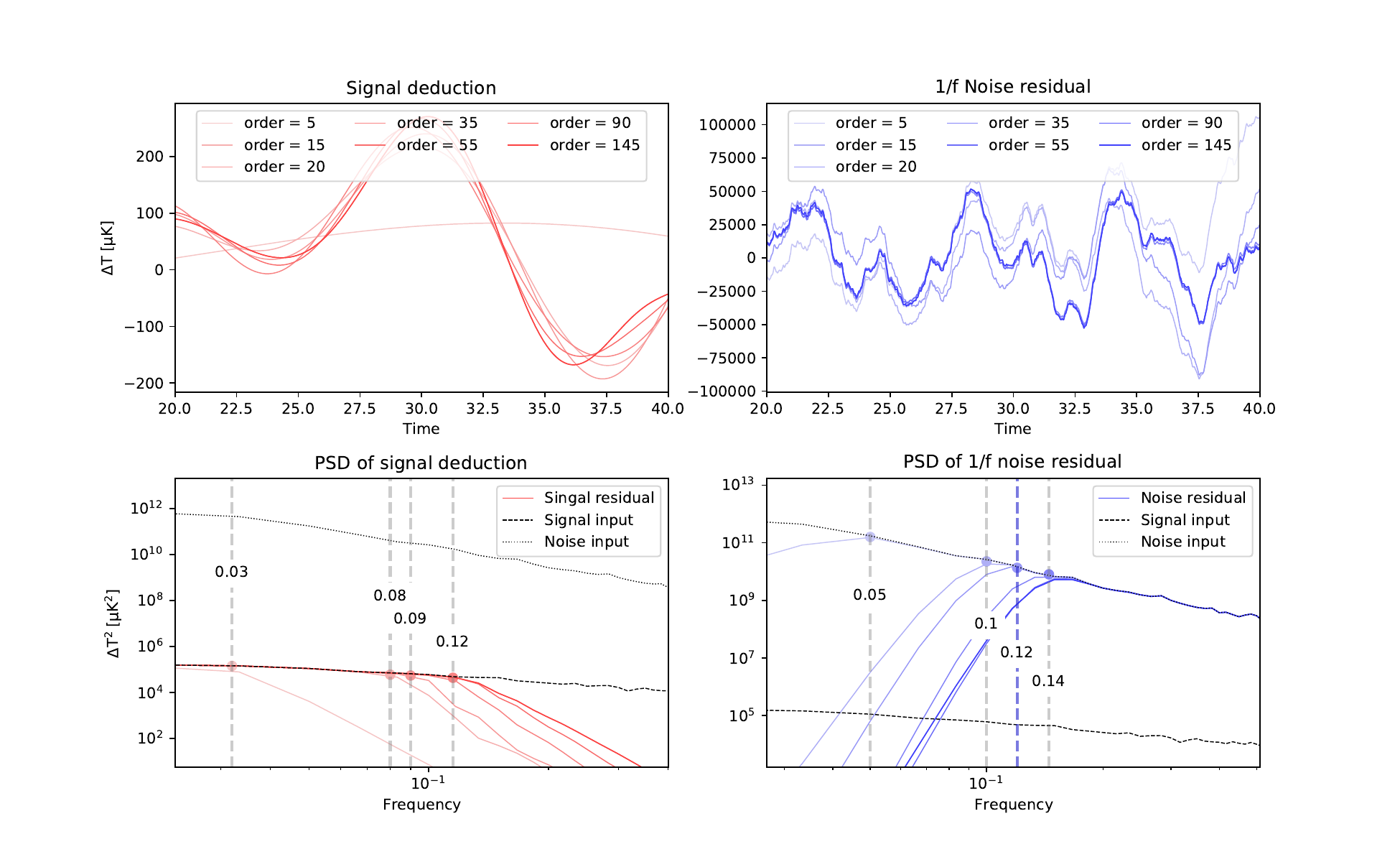}
    \end{center}
    \caption{The order test for low noise level TOD. Upper panel is the fitting curves for signal and $1/f$ noise under order from 5 to 145; lower panel is the corresponding PSD for both components. The blue vertical dashed line labels the corner frequency of 0.12.}\label{fig:poly_n_high}
\end{figure}

To perform polynomial fitting is to adopt eq.(\ref{EQ_polyfit}) on the TOD data, and the order of polynomial function depends on the featured modes that needed to be removed. We evaluated the order of the polynomials required for TOD filtering at different noise levels. The results for low noise level and high noise level are shown in Figure \ref{fig:poly_n_low} and Figure \ref{fig:poly_n_high}, respectively. The upper left subplots of both two figures are the result of direct polynomial fitting to the signal, which represents the filtering damage to the signal after filtering. The upper right plots are the $1/f$ residual noise after filtering. We use the red group of lines and the blue group of lines to represent the signal impairment and the noise residual, respectively, and the darker the color represents the higher order of the polynomial. The lower plots of Figure \ref{fig:poly_n_low} and Figure \ref{fig:poly_n_high} are the PSD of signal deduction and $1/f$ noise residual for different polynomial orders, respectively. 
The overall results show that the filter efficiency of polynomial fitting to the $1/f$ noise increases much less than the increase of polynomial fitting orders. This means that with the increase of order, it will become more and more difficult to fit the smaller features in TOD. Therefore, we must balance the order and the calculation cost. Considering all the factors mentioned above, we finally choose the order of 15 and 20 for low and high $1/f$ noise level,  respectively, and the corresponding corner frequencies---connecting the subtracted noise frequency domain and the noise residual frequency domain---are 0.11 and 0.12 for low and high $1/f$ noise level, respectively, which are labeled in blue vertical dashed lines in the right lower panel of both figures.


% \iffalse
\subsubsection{High-pass filter on TOD}

For each ring, the TOD was Fourier transformed to frequency domain, and the Butterworth high-pass filter coefficients was constructed with order of 5 and critical frequencies of 0.9 Hz (low noise level) or 2 Hz (high noise level), the Fourier coefficients of TOD were multiplied by filter coefficients to get the filtered coefficients, which are finally inverse Fourier transformed to the time domain, and the high-pass filtered TOD was obtained.  

\subsubsection{Wiener filter on TOD}

Like what we did in high-pass filter operation, the Wiener filter is also done in the frequency domain, and the filter coefficients is $W(f)=1/(1+\rho^{-1}(f))$ derived in section \ref{wiener}, then the filtered coefficients are calculated using the equation \ref{eq:wf}, finally the Wiener filtered TOD was obtained from the inverse transformation of filtered coefficients.

\subsection{Map making with TOD}
After we get the filtered TOD, the following step is to project the TOD onto map. The usual model for observed TOD is given by the summation of signal and noise:
\begin{equation}\label{optTOD}
\bm{d} = \bm{A}\bm{x}+\bm{n},
\end{equation}
where $\bm{d}$ is a vector which represents the observed TOD with length of number of samples, $\bm{x}$ is a vector consist by pixelized sky signal, $\bm{n}$ is the total noise including atmosphere, instrument, etc., and $\bm{A}$ is the pointing matrix. The unbiased solution of equation \ref{optTOD} is :
\begin{equation}\label{optEQ}
    \tilde{\bm{x}} =(\bm{A}^t \bm{W} \bm{A})^{-1} \bm{A}^t \bm{W} \bm{d},
\end{equation}
where the matrix $\bm{W}$ can be any positive definite matrix. In this paper we choose two different $\bm{W}$ to do the map making:

\begin{enumerate}
    \item \textbf{Naive Mapmaking:} the $\bm{W}$ is taken as a diagonal matrix, where the diagonal elements are the weights calculated from the inverse variance among one scan ring data, which means that for each pixel, where all the samples falling into are weighted co-added to give the final sky signal estimator.

    \item \textbf{Minimum Variance (MV) Mapmaking:} the $\bm{W}$ is taken as : $\bm{W} = \bm{N}^{-1}$, where the $\bm{N}$ is the noise covariance in time-domain, this estimator corresponds to the minimum variance or maximum likelihood solution. In this paper, we construct the $\bm{N}$ from the noise model after different filter operations and a detailed process of constructing $\bm{N}$ is illustrated in Appendix \ref{pcg}.
\end{enumerate}

%% file: results.tex
\label{sec:results}
In this section, we summarize the results of applying three filters to remove atmospheric $1/f$ noise.\footnote{Here we focus on the efficiency of filtering out 1/f-type noise within the atmospheric radiation. In this study we do not focus primarily on the filter's effect on the cancellation of atmospheric white noise, which can be reduced in the data by averaging of the TOD.} We compare the filtering results at two different noise levels of atmospheric emission, as defined in  subsection \ref{sec:atm_emis}, from several aspects. 
Firstly, in the TOD domain, we analyze the efficiency of each filters and compare their abilities using the criteria of the Root of Mean Square Error (RMSE) parameter estimated after the filtering process in subsectin \ref{sec:rmse}. Next, in subsection \ref{sec:analysis on naive mmk} we evaluate the effect of each filter in the map domain by estimating the power spectrum of the map after performing the filtering operation, including the signal power spectrum, the residual noise power spectrum, and comparing it with the input power spectrum. We estimate how the filters cut off the signals, as well as the atmospheric noise residual after performing the filters during the overall data processing. 
Finally, we show in subsection \ref{sec: pcg mmk} the results based on the optimized mapping method.

\subsection{Analysis at the level of TOD}\label{sec:rmse}

Filtering operations are typically carried out on TOD series, allowing us to assess the performance of a filter by comparing the effects on the signal and noise components in the TOD data before and after filtering.

To quantify the effectiveness of a filter, we define the Root of Mean Square Error (RMSE) parameter as follows: 
\begin{equation}
    E=\left( \sum^{N}_{i=0}(\hat x_i-x)^2/N\right) ^{\frac{1}{2}}.
\end{equation}
Here, $\hat x_i$ is the $i^{th}$ estimated value of the desired signal, and $x$ is the input simulated data. The RMSE provides the squared average of the differences between the input and output signals before and after filtering of the $N$ simulated data. A smaller RMSE value indicates that the filter output is closer to the desired signal.

\begin{figure}[bthp]
    \begin{center}
        \includegraphics[width=0.6\textwidth]{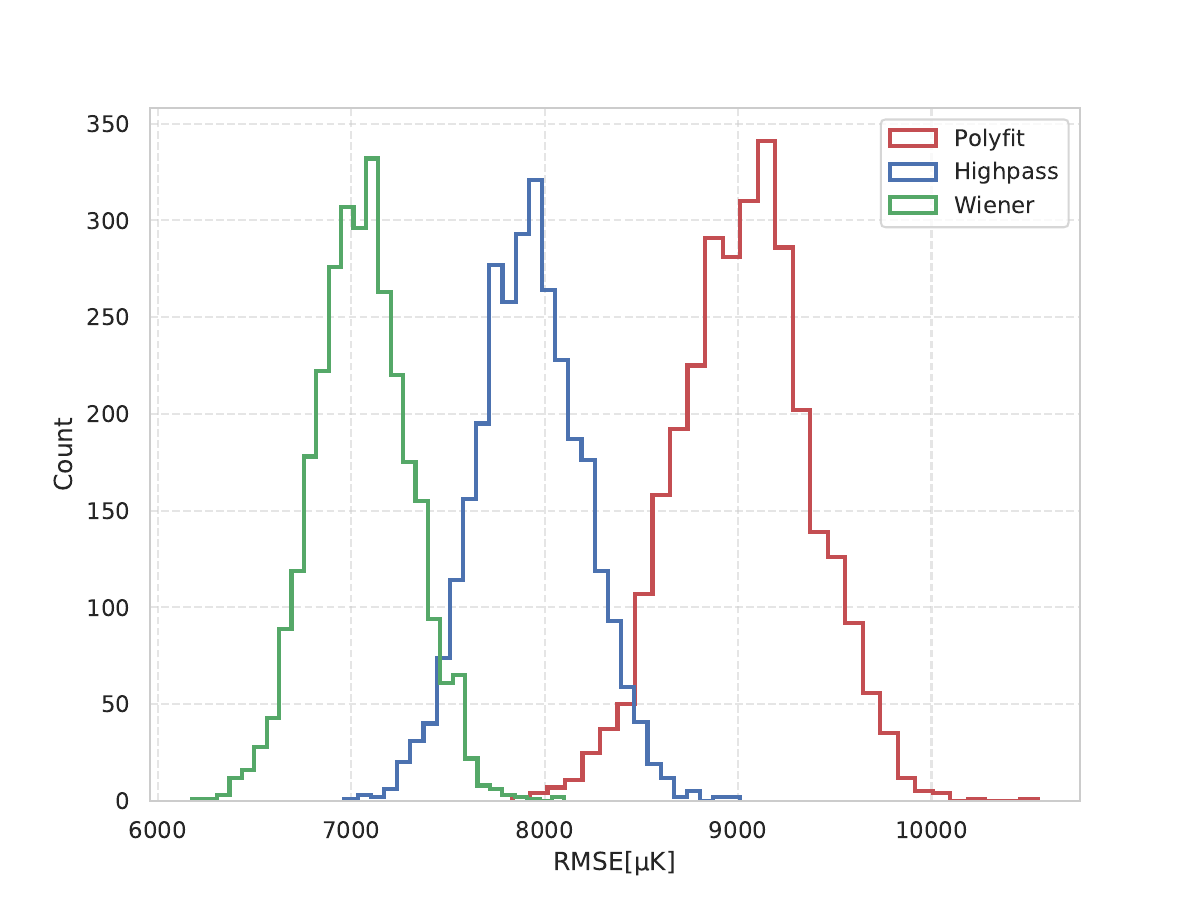}
    \end{center}
    \caption{The RMSE of three filters for low noise level. The red, blue, green lines are for the RMSE results at TOD domain by polynomial fitting, high-pass filter and Wiener filter, respectively.}\label{fig:rmse_low}
\end{figure}

\begin{figure}[bthp]
    \begin{center}
        \includegraphics[width=1\textwidth]{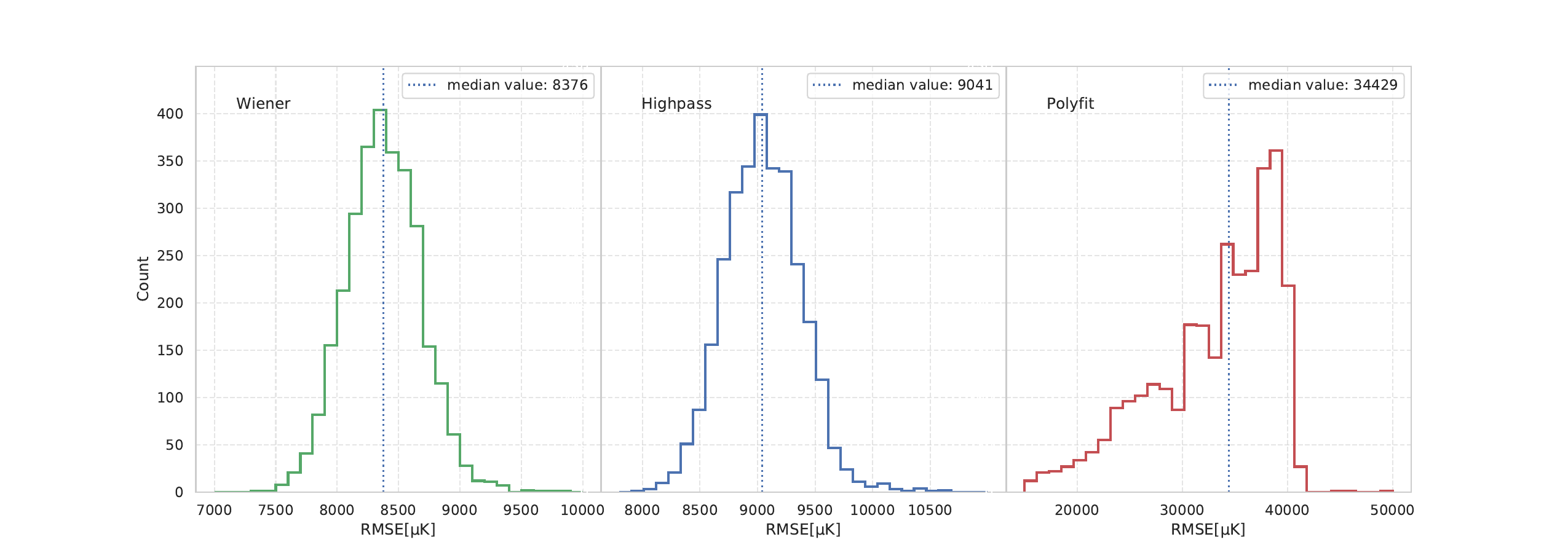}
    \end{center}
    \caption{The RMSE of three filters for high noise level. The median values of RMSE for three filters are marked with vertical dotted lines.}\label{fig:rmse_high}
\end{figure}

Figure \ref{fig:rmse_low} and Figure \ref{fig:rmse_high} show the RMSE results obtained by applying the three filters to the simulated TOD for low and high atmospheric noise levels, respectively. The results indicate that the polynomial fitting filter has the highest RMSE value for both low and high noise levels, followed by the Wiener filter, while the high-pass filter has the lowest RMSE value. In both cases, the RMSE value increases with the amount of atmospheric emission in the TOD. This can be attributed to the high level of white noise, which reduces the filtering efficiency, and the presence of large $1/f$ noise, which increases the residuals. The increase in RMSE values for polynomial fitting is more significant than for the high-pass and Wiener filters, indicating that the performance of the polynomial fitting filter decreases significantly as the noise level increases.

\begin{figure}[bthp]
    \begin{center}
        \includegraphics[width=0.8\textwidth]{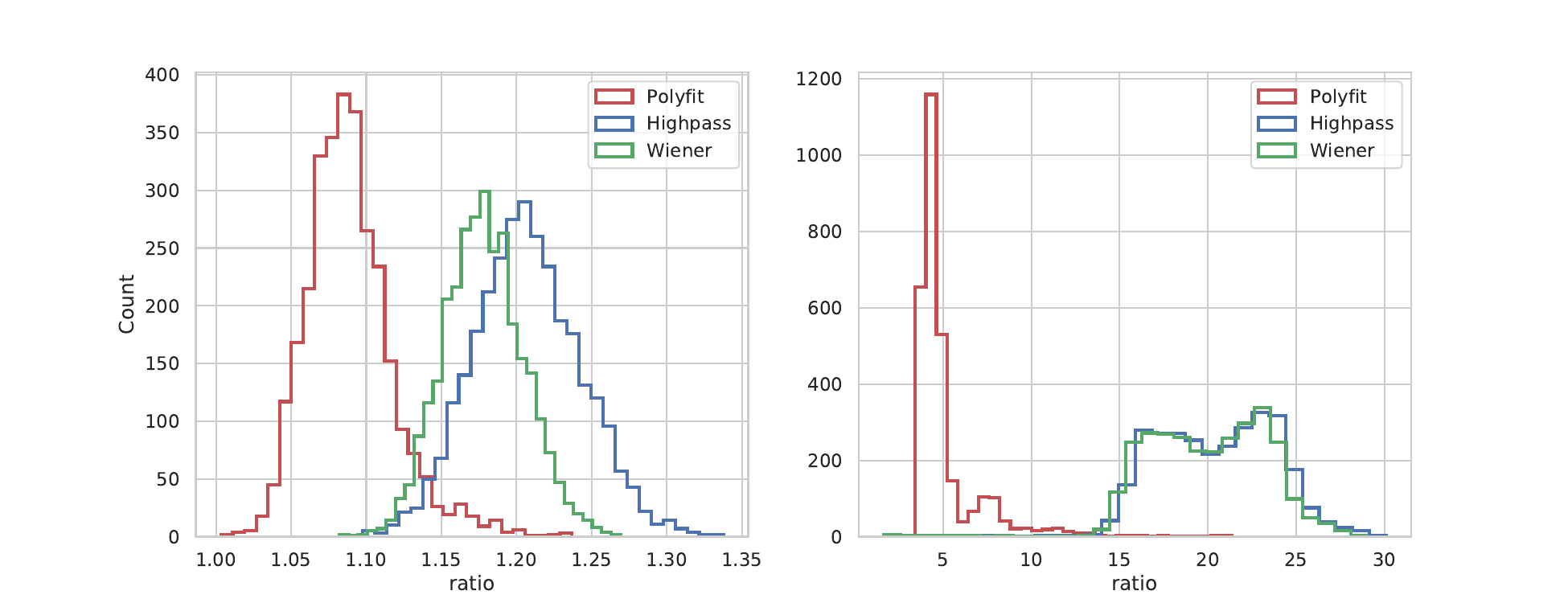}
    \end{center}
    \caption{The distributions of ratio for low atmospheric noise case (left) and high atmospheric noise (right) case. }\label{fig:snr}
\end{figure}

In order to compare the filtering efficiency more intuitively on the TOD domain, we have specifically calculated the ratio of the RMSE of the input TOD to the output TOD, which gives a more direct indication of the filtering efficiency. The ratio is defined as follows:
\begin{equation}
    \mathrm{ratio} = \frac{\mathrm{RMSE(input)}}{\mathrm{RMSE(output)}}~.
\end{equation}
 The distributions of the ratio for the low noise case and high noise case are shown in Figure \ref{fig:snr}, with the polynomial fitting having the significantly smaller values compared to the other two. Here the ratio characterizes the procedure of the atmospheric noise reduction of the TOD data after applying the filters, so that usually all ratio values should be greater than one. The results show that all three filters can improve the data quality more or less, and high-pass filter and Wiener filter perform better. Especially when the atmosphere noise is relatively high, on average they tend to reduce the RMSE by a factor of 20, which is more than four times better than the polynomial fitting.

\subsection{Analysis at the map domain}\label{sec:naive mmk}\label{sec:analysis on naive mmk}
The data processing at the TOD level is not the final output of CMB experiments. Instead, the estimation of the CMB angular power spectrum, along with the estimation of its variance, is the final quantity of interest. Therefore, we proceed to map the filtered TOD and examine the impact of the filters on the angular power spectrum at the map level. Our focus is on temperature maps, and we do not consider polarization in this study.

\subsubsection{Map differences after filtering}

\begin{figure}[bthp]
    \begin{center}
        \includegraphics[width=0.9\textwidth]{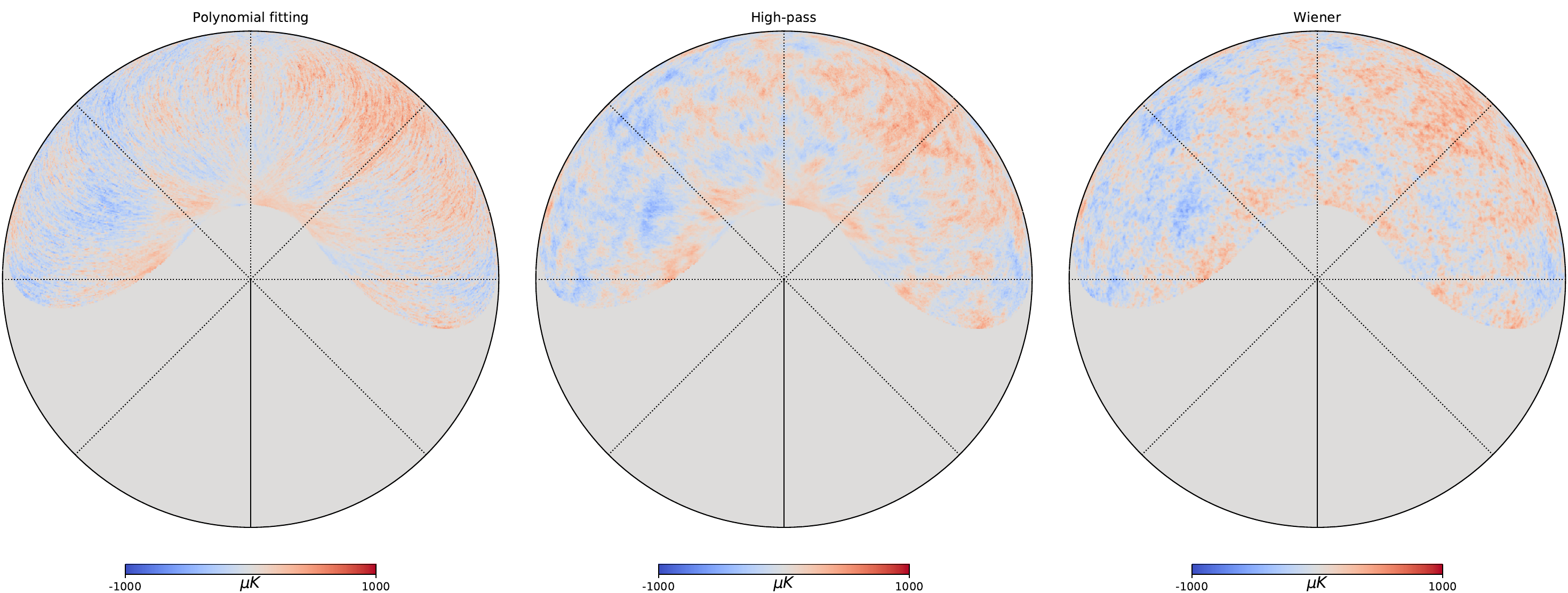}
    \end{center}
    \caption{The difference between the output and the input after filtering at low noise level. For polynomial fitting (left), high-pass filter (centre) and Wiener filter (right) respectively. }\label{fig:mapdifflow}
\end{figure}
\begin{figure}[bthp]
    \begin{center}
        \includegraphics[width=0.9\textwidth]{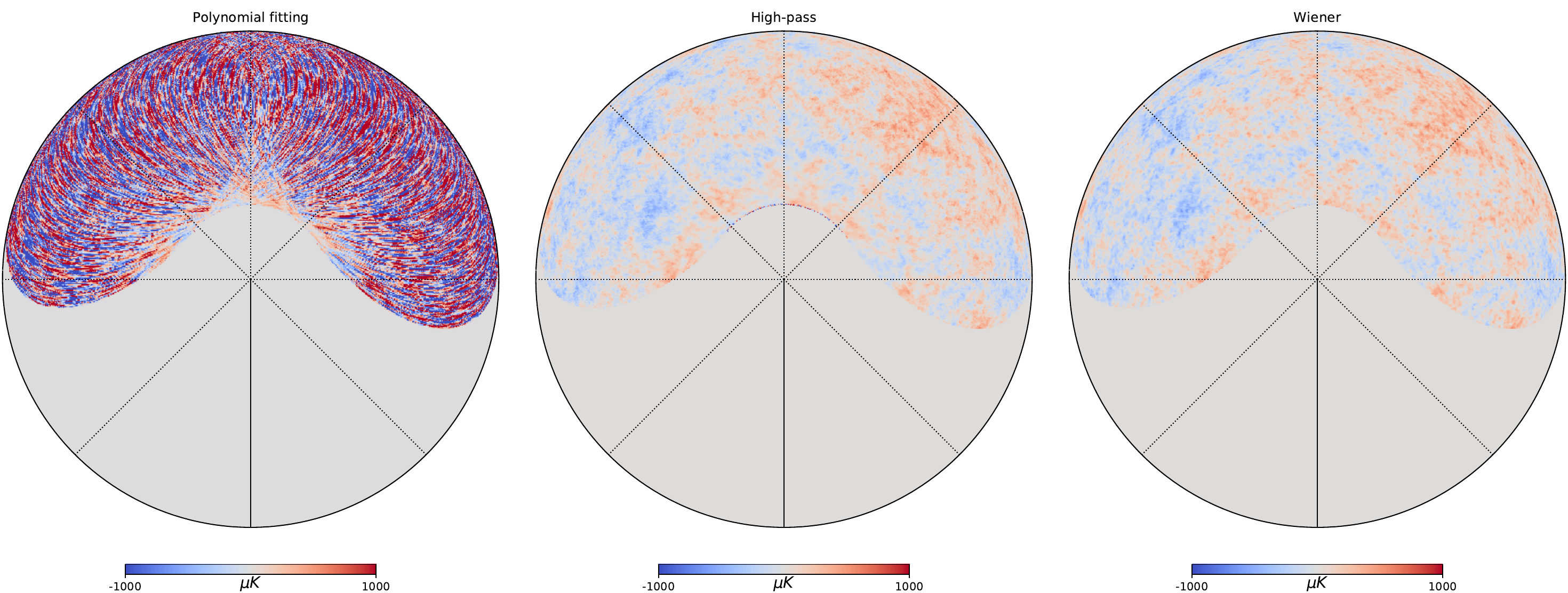}
    \end{center}
    \caption{The difference between the output and the input after filtering at high noise level. For polynomial filter (left), high-pass filter (centre) and Wiener filter (right) respectively.}\label{fig:mapdiffhigh}
\end{figure}
We performed mapmaking by projecting the filtered TOD onto maps, and analyzed the effects of the filters at the map level. In this subsection, we present a preliminary analysis based on the map differences of the three filters. 

Figure \ref{fig:mapdifflow} and Figure \ref{fig:mapdiffhigh} display the difference between the input data and the filtered output data using the naive mapmaking method, at low and high noise levels, respectively. These differences mainly represent the noise residual, but also include some of the signal variations. The left panel of Figure \ref{fig:mapdifflow} and Figure \ref{fig:mapdiffhigh} show large-scale streak structures in the map obtained from the polynomial fitting, indicating the leakage of noise residuals at large scales. The high-pass filter results (middle panel) display noise residuals on large scales, but with much lower magnitudes compared to the polynomial fitting. The Wiener filter (right panel) results do not show a significant deviation from the particular structure of the CMB temperature fluctuations on the map. As the noise level increases, the noise residuals after filtering also increase significantly, as shown in Figure \ref{fig:mapdiffhigh}, while the polynomial fitting yields higher noise residual values compared to the high-pass and Wiener filters. The noise residuals obtained from the Wiener filter appear slightly more uniform and have smaller amplitudes compared to those obtained from the high-pass filter.

\subsubsection{Power spectrum of noise residuals and signal after filtering}

The main objective of the filters is to remove the $1/f$ noise primarily contributed by the atmosphere, which is mostly present at low frequencies. We calculate the power spectrum of the residual noise and compare the filtering efficiency. We also estimate the power spectrum of the signal before and after filtering to derive the suppression effects of the filtering process.

\begin{figure}[bthp]
    \begin{center}
        \includegraphics[width=1\textwidth]{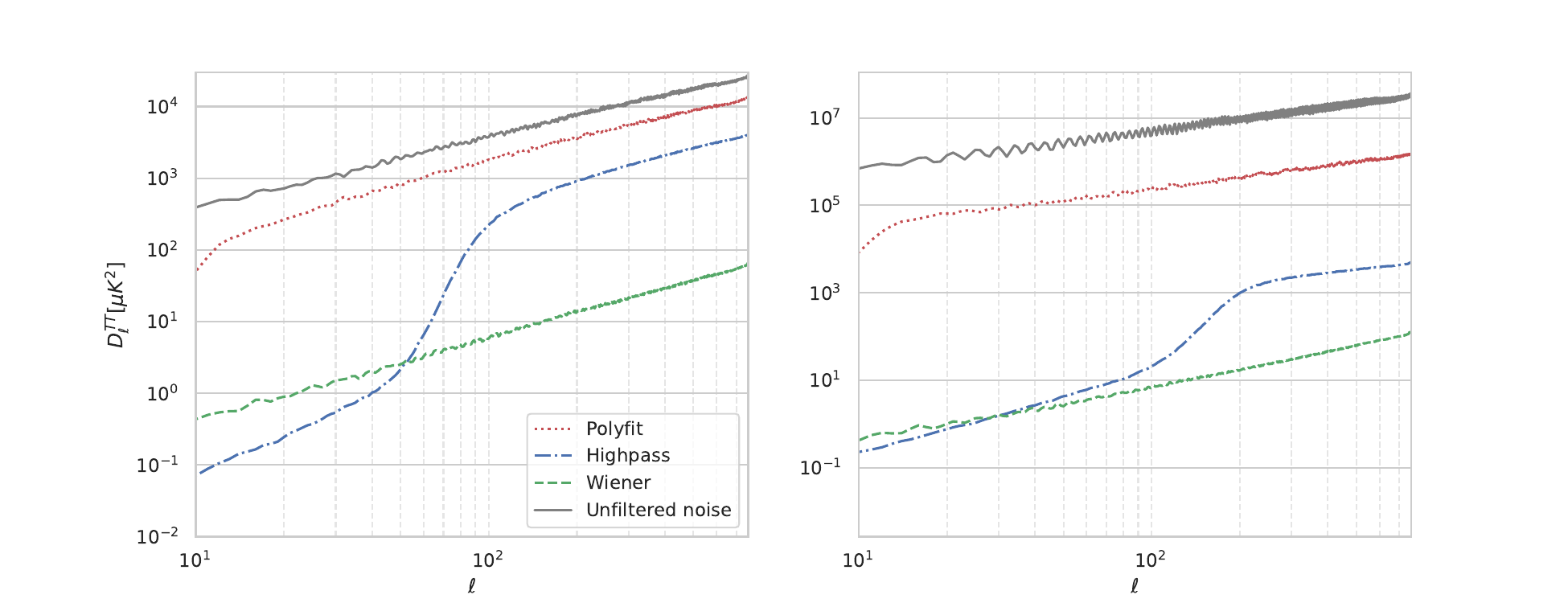}
    \end{center}
    \caption{The angular power spectrum of noise residuals for three filters. The gray solid line represents the input noise power spectrum without any filter operation, the red dotted line, green dashed line, and blue dot-dashed line correspond to the residual noise power spectrum obtained by polynomial fitting, high-pass filter, and winner filter, respectively. The left and right panels show the results for low and high noise levels, respectively.}\label{fig:noise_res}
\end{figure}

Figure \ref{fig:noise_res} shows the residual noise power spectrum for the two different noise levels in the left and right panels, respectively. At low noise level, the high-pass and Wiener filters effectively reduce the noise spectrum by about three to four orders of magnitude at large scales ($l<100$), while the polynomial fitting is more modest and mainly cuts off $1/f$ noise on very large scales of $l<11$ with a significantly smaller suppression order. At high noise level, similar results are obtained for the three filters, with the Wiener and high-pass filters suppressing the noise spectrum by five to six orders of magnitude on very large scales, while the polynomial fitting is much less effective. As the noise level increases, the efficiency of the Wiener and high-pass filters increases significantly. However, for the polynomial fitting, although its order is increased from 15 to 20, the noise reduction efficiency does not reach the level of the high-pass and Wiener filters, and the residual amount of noise is relatively high.

\begin{figure}[bthp]
    \begin{center}
        \includegraphics[width=1\textwidth]{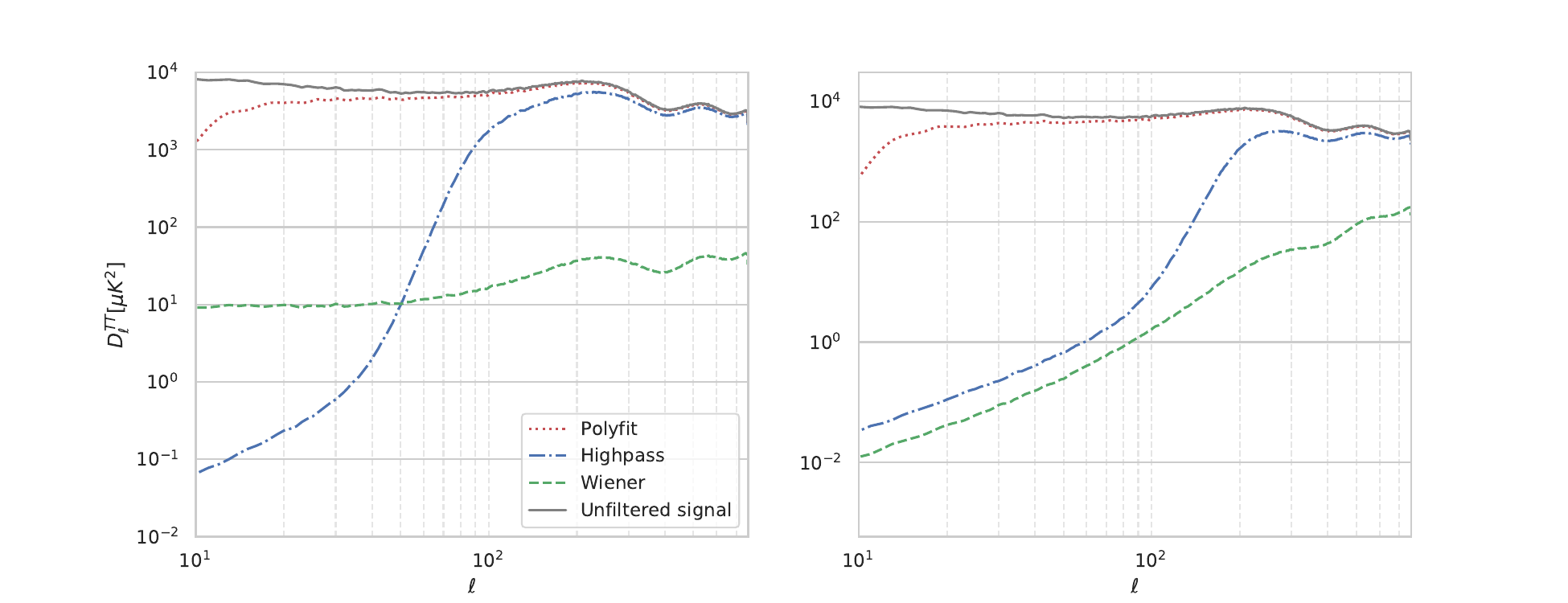}
    \end{center}
    \caption{The angular power spectrum of signal obtained by applying three filters. The gray solid line represents the input signal power spectrum without any filter operation, the red dotted line, green dashed line, and blue dot-dashed line correspond to the residual signal power spectrum obtained by polynomial fitting, high-pass filter, and winner filter, respectively. The left and right panels show the results for low and high noise levels, respectively.}\label{fig:snl_res}
\end{figure}

Since TOD is a linear combination of the signal and noise, we expect that filter has a similar effect on the input signal as the noise. Therefore, it is expected that the filter would suppress the signal while reducing noise. The fidelity of the signal given by the three filters are presented in Figure \ref{fig:snl_res}, in which the suppression effect on signal can be seen obviously, and they show the similar trend for both low and high noise levels. The numerical calculations indicate that the polynomial fitting causes the least distortion to the signal spectrum. In contrast, the high-pass filter results in significant suppression of the signal on large scale ($l<100$) for low noise level and $l<200$ for high noise level. The Wiener filter, unlike the polynomial and high-pass filters, suppresses the signal at almost all scales.
 
\subsubsection{Recovery of filtered signal}
Filtering typically leads to a loss of signal, and the suppression factor of the signal caused by filtering can be estimated statistically through multiple simulations of the filtering operation. The signal can be recovered by dividing by the suppression factor. In this study, we implemented filtering on 50 maps and estimated the average suppression factor for these 50 realizations. Using the suppression factors, we can recover the signal that were suppressed during the filtering process.

During the process of recovering the signal power spectrum, the suppression factor also amplifies the residual noise. As a result, the final recovered noise level is higher than the residual noise in the absence of recovery. Since signal recovery is a necessary process during CMB data analysis, we compare the final noise levels based on the results after recovery.

\begin{figure}[bthp]
    \begin{center}
        \includegraphics[width=1\textwidth]{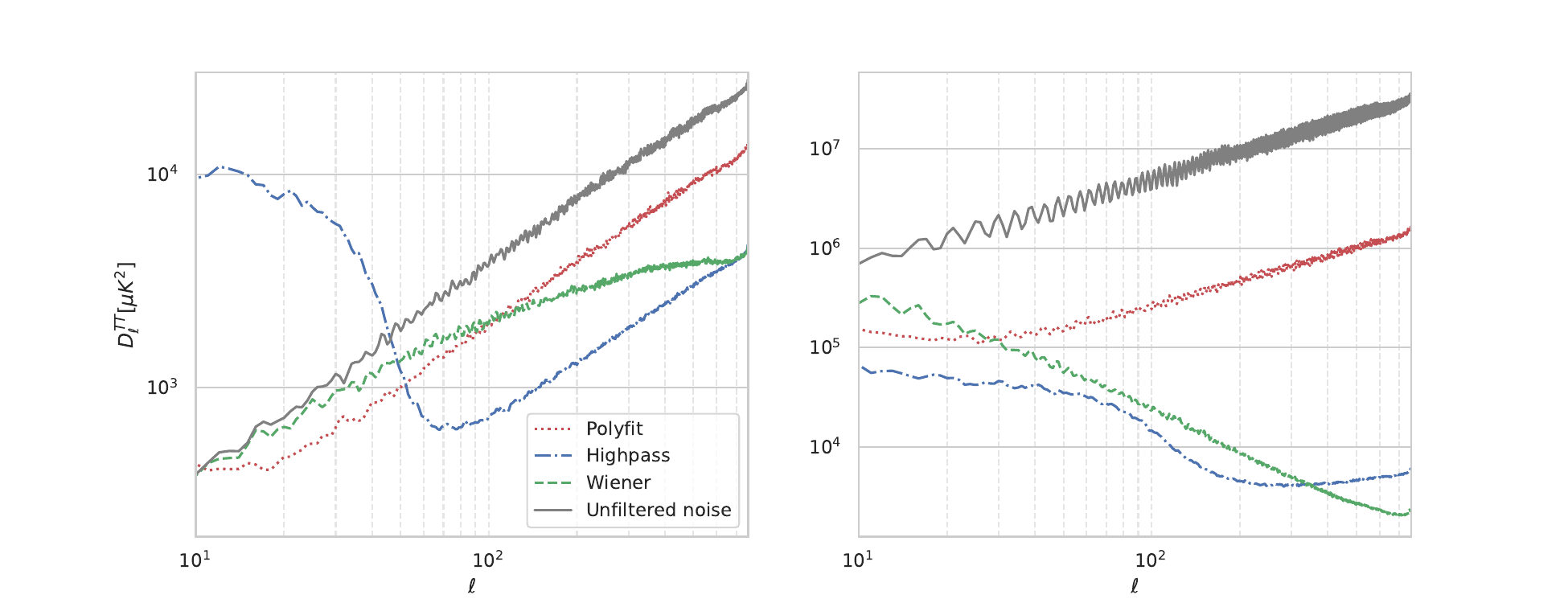}
    \end{center}
    \caption{The angular power spectrum of recovered noise. The gray solid line is the input noise power spectrum without any filter operation, and red dotted line, green dashed line, and blue dot-dashed line represent the recovered noise power spectrum of polynomial fitting, high-pass filter and Wiener filter, respectively. The left and right panels are for low and high noise level, respectively.}\label{fig:noise_rec}
\end{figure}

We present a comparison of the power spectrum of the modified noise residuals for the three filters in Figure \ref{fig:noise_rec}. The results show that for low noise level, the modified noise spectrum of the Wiener filter and polynomial fitting are lower than that of the high-pass filter at large scales ($\ell<50$). For high noise level, the polynomial fitting still results in higher noise spectrum than the other two filters across most scale ranges. Notably, the high-pass filter produces higher noise spectrum at large scale than the unfiltered input noise after performing signal recovery.  At small scales, the Wiener filter and high-pass filter exhibit lower noise spectrum than polynomial fitting for both low and high noise levels.

\subsubsection{Standard deviation of power spectrum}\label{sec:5.2.4}
After recovering operation, the signal is corrected to the same level as the input value for three filters.  To evaluate how data filtering affects the accuracy of CMB measurements, we present covariance statistics for the corrected power spectrum.

\begin{figure}[bthp]
    \begin{center}
        \includegraphics[width=1\textwidth]{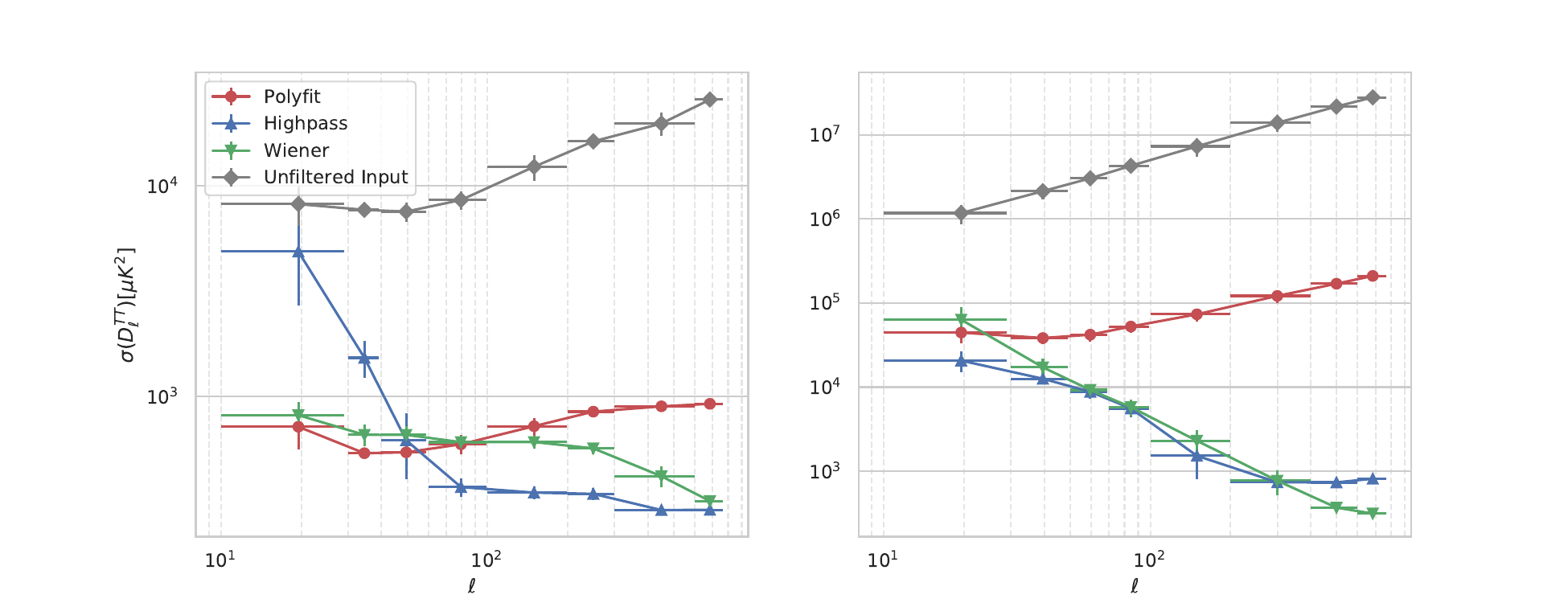}
    \end{center}
    \caption{Standard deviation of binned power spectrum after filtering. Red, blue, green lines stand for the results after applying the polynomial fitting, high-pass filter and Wiener filter, respectively. Dots and Y-axis error bars represent the mean value and one sigma error of the $\sigma(D_{\ell}^{\mathrm{TT}})$s in the bin. X-axis error bars show the multipole range. Left panel and right panel are for low and high noise level, respectively.}\label{fig:std}
\end{figure}

The accuracy of the power spectrum measurement increases as the standard deviation of the power spectrum decreases. Figure \ref{fig:std} presents the standard deviation of the power spectrum obtained after data filtering and recovery. At low noise level (left panel), the polynomial fitting and Wiener filter produce similar levels of standard deviation for the power spectrum, while the high-pass filter yields much higher values at larger scales ($l<50$), where the standard deviation can be one order of magnitude higher than the values obtained from the polynomial fitting and Wiener filter. However, at smaller scales, the high-pass filter performs well, resulting in smaller standard deviations than the polynomial fitting and Wiener filter.

In the case of high noise level (right panel), the standard deviation values of the power spectrum given by the three filters are comparable in magnitude at very large scales (above $l<30$). At other scales, the results of the high-pass filter and the Wiener filter are comparable, and both produce much better results than the standard deviation value given by the polynomial fitting.

\subsection{Analysis based on optimal map making}\label{sec: pcg mmk}

\begin{figure}[bthp]
    \begin{center}
        \includegraphics[width=1\textwidth]{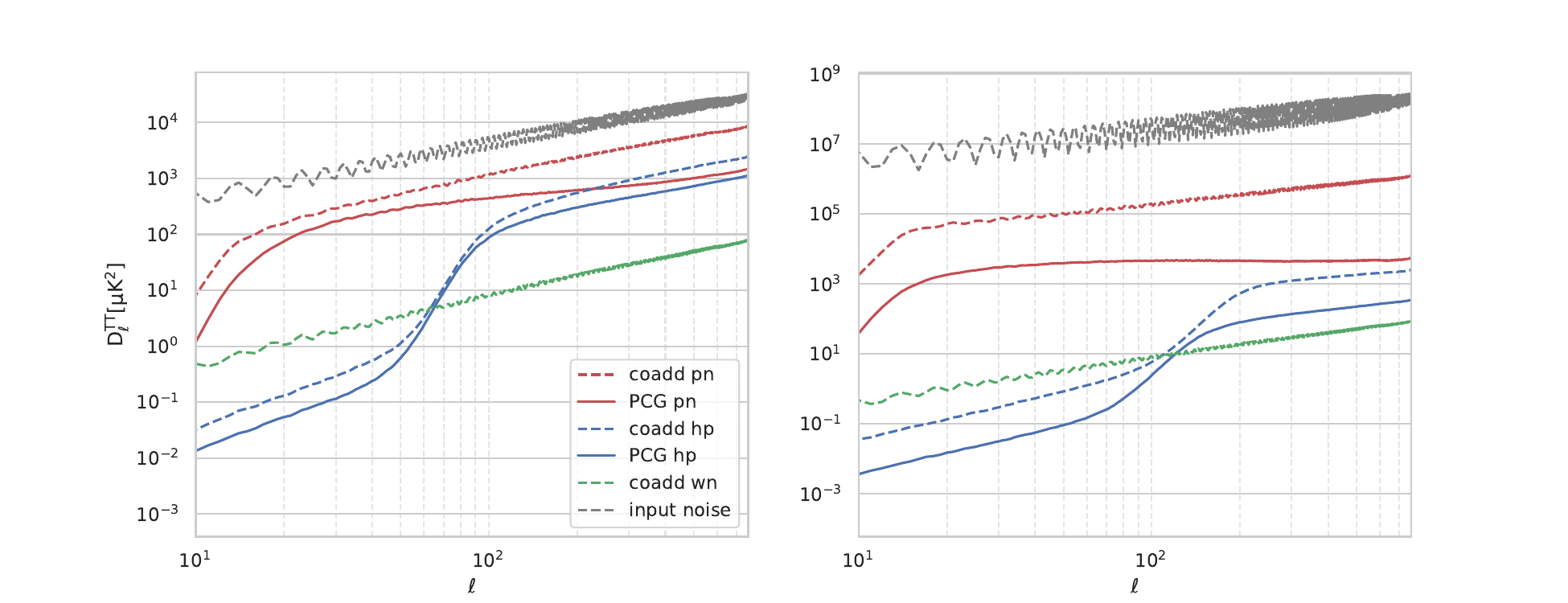}
    \end{center}
    \caption{The power spectrum of noise residual by three filters, using an optimised map making method (labeled with `PCG' and represented by the solid lines) and a naive map making method (labeled with ``coadd" and represented by the dashed lines). The left and right panels represent the results obtained at low and high noise level, respectively. The red, blue and green lines represent after polynomial fitting, high-pass, Wiener filter, respectively, and the gray lines are the input noise.}\label{fig:pcgnoiseres}
\end{figure}

\begin{figure}[bthp]
    \begin{center}
        \includegraphics[width=1\textwidth]{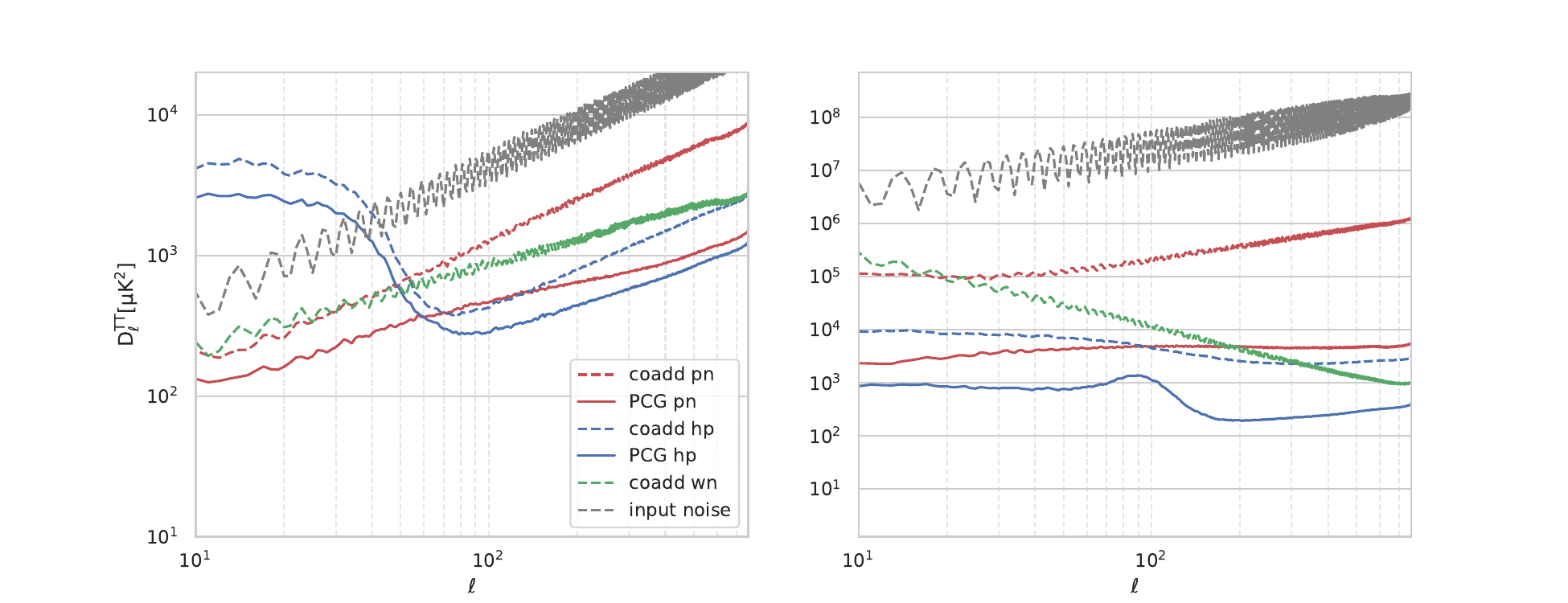}
    \end{center}
    \caption{The power spectrum of noise recovery,  using an optimised map making method (labeled with `PCG' and represented by the solid lines) and a naive map making method (labeled with ``coadd" and represented by the dashed lines). The solid lines and dashed lines represent the results obtained by optimised map making method and naive map making method, respectively. The left and right panels represent the results obtained at low and high noise level, respectively. The red, blue, green, and gray lines represent polynomial fitting, high-pass, Wiener filter, and noise input, respectively. }\label{fig:pcgnoiserec}
\end{figure}

In this section, we investigated the noise suppression effect of the map making methods. To compare the results of the naive map making method with an optimized approach, we calculated the noise power spectrum of the filtered data using both methods and compared the results. The results are shown in Figure \ref{fig:pcgnoiseres} and Figure \ref{fig:pcgnoiserec}, where the dashed lines represent the noise power spectrum obtained using the naive mapping method, and the solid lines show the optimized mapping results labeled with `PCG'.

In Figure \ref{fig:pcgnoiseres}, we present the residual noise spectrum for two map-making methods.  
The left panel of Figure \ref{fig:pcgnoiseres} shows the results obtained at low noise level for the polynomial fitting and high-pass filter using both the naive and optimized mapping methods. The optimized mapping method does help to reduce the noise residual power spectrum compared to the naive mapping method. In the right panel, which shows high noise level results, the optimized mapping method performs even better. Specifically, for different multipole values, the polynomial fitting shows an improvement of about $20\thicksim 200$ times, while the high-pass filter shows an improvement of about 10 times. The main reason for this difference is that the polynomial fitting operation is unable to remove the $1/f$ noise component, whereas the high-pass filter is more effective at reducing this noise because it directly truncates the TOD low frequency components.

It's worth noting that we have not implemented an optimized mapping approach for the Wiener-filtered TOD. This is primarily because the noise covariance matrix of the Wiener-filtered TOD is diagonal, which means the optimized map making is equivalent to the naive map making. Appendix \ref{app:min var mmk} provides a theoretical analysis to explain this point.

Similar to Figure \ref{fig:noise_rec}, we compare the noise spectrum after recovering the signal loss for different filters in Figure \ref{fig:pcgnoiserec}. We confirm that naive mapping methods give higher noise residual power spectrum than the optimized mappings even after the signal recovery, and the high-pass filter gives more residuals for $\ell<100$, mainly because it uses direct truncation, resulting in a larger recovery factor.

In summary, the optimized mapping method can help to reduce the noise residual as long as we have a good knowledge of the noise covariance matrix. However, for actual observations whose atmospheric noise varies with time in an unpredictable way, the total noise covariance is hard to obtain; thus, the optimized mapping method should be used with care, which will be studied in a following work.

%% file: conclusion.tex
\label{sec:conclusions}
In this work, we compared the polynomial fitting, the high-pass filter, and the Wiener filter as three alternative filtering techniques for removing atmospheric noise from CMB measurements. To produce time stream data of a focal plane detector array with a circular scan strategy, a realistic simulation of the ground-based experiment is performed. Three distinct filtering procedures are then carried out on the simulated data, and the filtered data streams are projected onto pixel domain maps and compared with the residual noise, demonstrating that the residual noise for polynomial fitting is significant at low frequencies (large angular scales) and also leaks into small scales during the map making, whereas the high-pass and Wiener filters do not produce significant leakages at the map level.

We analyse end-to-end data and run time domain simulations to assess the variance of final signal power spectra with various filters and atmospheric noise levels. The findings demonstrate that all filters operate well at low atmospheric noise levels, with the exception that the high-pass filter produces greater noise residuals at large angular scales. In the situation of a high atmospheric noise level, the polynomial filter performs worse than the other two at both the medium and small angular scales. As a result, the Wiener filter performs the best overall, and the polynomial filter needs to be utilised with care when there is a high atmospheric noise.

One additional issue that needs to be brought up is, unlike the Wiener filter, the polynomial filter and the high-pass filter do not require prior knowledge of the signal and noise, particularly the spectral shapes of the CMB signal and the atmospheric noise. However, we have many measurements of the CMB and Galactic foreground power spectra, as well as a large number of atmospheric radiation spectrum measurements, which provides sufficient prior information to support the Wiener filtering operation. Meanwhile, the Wiener filter is not sensitive to small fluctuations of the prior information; thus, the requirement on the prior information is not a critical issue.

%% file: appendix.tex
\appendix

\section{The minimum variance and naive map making methods}
\label{app:min var mmk}

\subsection{Estimation from the TOD domain}
The CMB sky maps are produced from the time order data (TOD), which is called a map making. There are usually three ways: 1) Naive map making, which is to average the observations at each sky point with a necessary inverse-noise weighting. 2) Minimal variance map making without prior information of the CMB signal, which is solved in \citep{1997PhRvD..56.4514T}. 3) Minimal variance map making with prior information of the CMB signal, which is beyond this work. In this appendix, we make the point that cases 1) and 2) are really two distinct perspectives of the same issue; as a result, once our approaches have been tested in case 1), they are also applicable to case 2).

According to \citep{1997PhRvD..56.4514T}, the minimal variance map making in case 2) is based on the following equation:
\begin{align}\label{equapp:basic}
    \bm{d} = \bm{A}\bm{x} + \bm{n}_1,
\end{align}
where $\bm{d}$ is the time order data, $\bm{A}$ is the pointing matrix that converts the sky map $\bm{x}$ into the time order data, and $\bm{n}_1$ is the time-order noise, which usually contains both white and red noises, as well as atmosphere and ground noises (if the experiment is ground based). Therefore, the noise covariance matrix $\bm{N}_1=\langle \bm{n}_1\bm{n}_1^t \rangle$ is usually non-diagonal. The map making tries to get a minimal variance estimation of the sky map $\bm{x}$, and the answer is given by \citep{1997PhRvD..56.4514T} as
\begin{align}\label{equapp:tegmark solution}
    \bm{\Tilde{x}} = (\bm{A}^t\bm{N}_1^{-1}\bm{A})^{-1}\bm{A}^t\bm{N}_1^{-1}\bm{d}.
\end{align}
Now we prove that this solution is nothing but a different view of the naive map making, which is characterized as the well known equation $\bm{\Tilde{x}} = (\bm{A}^t\bm{A})^{-1}\bm{A}^t\bm{d}$.

The noise covariance matrix $\bm{N}_1=\langle \bm{n}_1\bm{n}_1^t \rangle$ is apparently real and symmetric; thus, it can always be diagonalized as
\begin{align}
    \bm{N}_1 = \bm{B}^t\bm{\lambda}\bm{B},
\end{align}
where $\bm{\lambda}$ is diagonal and all its diagonal elements are positive; and $\bm{B}$ is unitary. Therefore, we obtain
\begin{align}\label{equapp:diag noise}
    \bm{N}_1 = \langle \bm{n}_1\bm{n}_1^t \rangle = \bm{B}^t\bm{\lambda}^{1/2}\bm{I}\bm{\lambda}^{1/2}\bm{B} = \bm{B}^t\bm{\lambda}^{1/2}\langle\bm{n}_0\bm{n}_0^t\rangle\bm{\lambda}^{1/2}\bm{B},
\end{align}
where $\bm{n}_0$ represents a white noise component with standard normal distribution, whose covariance matrix is $\langle\bm{n}_0\bm{n}_0^t\rangle = \bm{I}$. Thus, a noise component $\bm{n}_1$ whose covariance matrix is non-diagonal can always be converted to standard white noise $\bm{n}_0$ by a unitary matrix whose rows are weighted by a diagonal matrix $\bm{\lambda}^{1/2}$.

Now we move back to eq.~(\ref{equapp:basic}) and substitute $\bm{n}_1 = \bm{B}^t\bm{\lambda}^{1/2}\bm{n}_0$ to get
\begin{align}
    \bm{d} &= \bm{A}\bm{x} + \bm{n}_1 = \bm{A}\bm{x} + \bm{B}^t\bm{\lambda}^{1/2}\bm{n}_0,
\end{align}
which gives
\begin{align}
    \bm{\lambda}^{-1/2}\bm{B}\bm{d} &= \bm{\lambda}^{-1/2}\bm{B}\bm{A}\bm{x} + \bm{n}_0.
\end{align}
Because $\bm{B}$ is unitary, it is safe to write $\bm{A}_1 = \bm{B}\bm{A}$, which indicates $\bm{A}$ and $\bm{A}_1$ are views of the same thing in different reference frames. According to eq~(\ref{equapp:tegmark solution}) and considering the fact that $\langle\bm{n}_0\bm{n}_0^t\rangle=\bm{I}$, the minimal variance solution of the above equation is simply
\begin{align}
    \tilde{\bm{x}} &= ((\bm{\lambda}^{-1/2}\bm{A}_1)^t(\bm{\lambda}^{-1/2}\bm{A}_1))^{-1}(\bm{\lambda}^{-1/2}\bm{A}_1)^t(\bm{\lambda}^{-1/2}\bm{B}\bm{d}) \\ \nonumber
    &= (\bm{A}_1^t\bm{\lambda}^{-1}\bm{A}_1)^{-1}\bm{A}_1^t\bm{\lambda}^{-1}\bm{B}\bm{d} \\ \nonumber 
    &= (\bm{A}_2^t\bm{A}_2)^{-1}\bm{A}_2^t(\bm{B}'\bm{d}),
\end{align}
where $\bm{A}_2=\bm{\lambda}^{-1/2}\bm{A}_1$ and $\bm{B}'=\bm{\lambda}^{-1/2}\bm{B}$. The above equation is exactly a naive map making with inverse-noise weighting (so data with higher noise level get lower weights). Therefore, the minimal variance map making and the naive map making are views of the same thing in different reference frames. This means in principle, if our approaches are tested with the naive map making, it should have no problem with the minimal variance map making given in \citep{1997PhRvD..56.4514T}.

\subsection{Estimation from the pixel domain}
The minimum variance estimation of the sky map from the TOD domain was discussed above.  However, this estimation is not always available, because the TOD domain noise covariance matrix is unstable in some cases, for example, when the TOD contain the atmospheric noise that is associated with the weather. Therefore, it is also preferable to estimate the sky map directly in the pixel domain, which means to left-multiply $\bm{A}^t$ to both sides of eq.~(\ref{equapp:basic}) in advance, so the estimation starts from the following:
\begin{align}\label{equapp:start from pix domain}
    \bm{A}^{t}\bm{d} = \bm{A}^t\bm{A}\bm{x} + \bm{A}^t\bm{n}_1,
\end{align}
where $\bm{A}^{t}\bm{d}$ is the coadded pixel domain sky map, and $\bm{A}^t\bm{n}_1$ is the coadded pixel domain noise. 

In this section, we point out the following fact: when we start from eq.~(\ref{equapp:start from pix domain}) to estimate the sky map, the minimal variance and naive estimations are identical, and both are completely unaffected by the property of noise $\bm{n}_1$: Because eq.~(\ref{equapp:start from pix domain}) and eq.~(\ref{equapp:basic}) have similar forms, the solution of eq.~(\ref{equapp:basic}) is also applicable to eq.~(\ref{equapp:start from pix domain}). Also note the fact that for simple coadding, $\bm{A}^t\bm{A}=\bm{\lambda}$ is a diagonal matrix, and we diagonalize the noise term as $\bm{A}^t\bm{n}_1=\bm{B}\bm{n}_0$ like we have done in eq.~(\ref{equapp:diag noise}) to get
\begin{align}
    \bm{A}^{t}\bm{d} &= \bm{\lambda}\bm{x} + \bm{B}\bm{n}_0 \\ \nonumber
    \bm{B}^{-1}\bm{A}^{t}\bm{d} &= \bm{B}^{-1}\bm{\lambda}\bm{x} + \bm{n}_0,
\end{align}
where $\bm{B}\bm{B}^t=\bm{N}$ is the pixel domain noise covariance matrix. Obviously, $\bm{N}^{-1} = \bm{B}^{-1,t}\bm{B}^{-1}$. Then the minimal variance solution to the above equation is
\begin{align}
    \bm{\Tilde{x}} &= ((\bm{B}^{-1}\bm{\lambda})^t(\bm{B}^{-1}\bm{\lambda}))^{-1} (\bm{B}^{-1}\bm{\lambda})^{t}\bm{B}^{-1}\bm{A}^{t}\bm{d} \\ \nonumber
    &= (\bm{\lambda}\bm{N}^{-1}\bm{\lambda})^{-1} \bm{
    \lambda}\bm{N}^{-1}\bm{A}^{t}\bm{d} \\ \nonumber
    &= \bm{\lambda}\bm{A}^t\bm{d},
\end{align}
which is identical to the naive map making and is free from the pixel domain noise property. 

Therefore, when we start from eq.~(\ref{equapp:start from pix domain}), the naive map making and the minimum variance map making (without prior knowledge of the desired signal) are identical. This fact is especially useful when the TOD domain noise is complicated and unstable, and the TOD domain noise covariance matrix is consequently unreliable: it enables us to keep the minimum variance condition as much as possible while utilising all available techniques (such as polynomial, high pass, and wiener filters) to minimise noise in the TOD domain and improve the sky map. Afterwards, the signal covariance matrix will be used to do the final sky map optimization in the pixel domain, which will be covered in a following work.

\section{Preconditioned Conjugated Gradient map making}
\label{pcg}
%\tred{To obtain the brute force solution of eq. (\ref{optEQ}) can be computationally expensive, especially for the process of solving $(\bm{A}^t\bm{N}^{-1}\bm{A})^{-1}$. This is because the inverse of a large matrix can require a significant amount of CPU time and memory to calculate, especially for sparse matrices commonly found in numerical methods. However, there are methods that can be used to reduce the computational cost of solving eq. (\ref{optEQ}).
%One such method is the preconditioned conjugate gradient (PCG) method, which is a widely used iterative method for solving large linear systems of equations. It is particularly useful for solving sparse linear systems, where most of the entries in the coefficient matrix are zero. The basic idea behind the PCG method is to combine the conjugate gradient (CG) method with a preconditioner that improves the convergence of the CG algorithm. The preconditioner is a matrix that approximates the inverse of the coefficient matrix and is used to transform the original linear system into an equivalent system that is easier to solve.
%Therefore, in order to reduce the computational cost of solving eq. (\ref{optEQ}), we applied the PCG method and rewrote it as:
%where $N$ is the noise covariance. This allows us to solve the linear system in a more efficient manner, while still achieving accurate results.}
As matrix $\bm{A}^t\bm{N}^{-1}\bm{A}$ in eq. (\ref{optEQ}) is too big to be inverted directly for large sky coverage, we choose the preconditioned conjugate gradient (PCG) method, which is a widely used iterative method for solving large linear systems of equations to solve the linear equation in eq (\ref{eq:least-sq}).  

\begin{equation} \label{eq:least-sq}
\bm{A}^t \bm{N}^{-1} \bm{A} \tilde{\bm{x}} = \bm{A}^t \bm{N}^{-1} \bm{d},
\end{equation}

During the PCG, we first introduce the preconditioned matrix $\bm{P}$ which is an approximate inverse of $\bm{A}^t\bm{N}^{-1}\bm{A}$ to make the convergence of iteration more quickly, and the eq. (\ref{eq:least-sq}) becomes a better-conditioned linear system:

\begin{equation} \label{eq:pcg}
\bm{P}\bm{A}^t \bm{N}^{-1} \bm{A} \tilde{\bm{x}} = \bm{P}\bm{A}^t \bm{N}^{-1} \bm{d},
\end{equation}

We summarize the PCG process as follows:
\begin{enumerate}
    \item 
        Building the pointing matrix $A$: This matrix was derived from the TOD simulation procedure, where we assumed that all the detectors shared the same pointing directions with the boresight direction of the virtual telescope.
    \item
        Building the unfiltered noise variance matrix $N_r$ for single ring: In subsection \ref{sec:atm_emis} we have modeled noise Power Spectrum (PS) with form $P(f)=\alpha f^{-\beta}$, here we plus it with white noise, which turns to $P(f)=\alpha f^{-\beta}+\sigma_n^2$. According to Parseval's Theorem, the auto-correlation function is the iFFT form of PS. Then, by toeplitzing the auto-correlation function we obtain the variance matrix $N_r$. 
    \item
        Obtaining the inverse of filtered noise variance matrix $N_{rf}^{-1}$: For polynomial fitting, the filter matrix can be written as $F=I-X(X^TX)^{-1}X^T$, where $X$ is a Vandermonde matrix. For high-pass filter, the filter matrix can be written as $F=\frac{1}{m}D^*HD$, where $D$ is the Fourier transform matrix and $H$ is a diagonal matrix with diagonal elements of filtering factors. We then obtained the inverse of filtered noise variance matrix $N_{rf}^{-1}=(FNF^T)^{-1}$. 
    \item
        Building the covariance matrix for a scan set: We marked the coefficient matrix of $\tilde{x}$ by $\mathcal{N}=A^tN^{-1}A$. We ignored the correlation between different rings in our study, so $\mathcal{N}$ could be easily realized by filling the diagonal with $N_{rf}$. As a block matrix, the inverse of $\mathcal{N}$ is the combination of the inverse of every filling matrix $N_{rf}$.
     \item   
        Running the code to obtain $\tilde{x}$: We input every filtered simulation TOD scan set to the PCG code, and each pixel of the final map was the arithmetic average of 400 night scan sets of PCG results at the corresponding pixel.
\end{enumerate}
By following these steps, we were able to use the PCG method to efficiently solve eq.  (\ref{optEQ}) and obtain accurate results.